# Thermodynamic anomalies and three distinct liquid-liquid transitions in warm dense liquid hydrogen

Hua Y. Geng[1,2,*], Q. Wu[1], Miriam Marqués[3], and Graeme J. Ackland[3]

[1]*National Key Laboratory of Shock Wave and Detonation Physics, Institute of Fluid Physics, CAEP; P.O.Box 919-102 Mianyang, Sichuan, P. R. China, 621900*

[2]*Center for Applied Physics and Technology, HEDPS, and College of Engineering, Peking University, Beijing 100871, China*

[3]*CSEC, SUPA, School of Physics and Astronomy, The University of Edinburgh, Edinburgh EH9 3JZ, United Kingdom*

## Abstract

The properties of hydrogen at high pressure have wide implications in astrophysics and high-pressure physics. Its phase change in the liquid is variously described as a metallization, $H_2$-dissociation, density discontinuity or plasma phase transition. It has been tacitly assumed that these phenomena coincide at a first-order liquid-liquid transition (LLT). In this work, the relevant pressure-temperature conditions are thoroughly explored with first-principles molecular dynamics. We show there is a large dependency on exchange-correlation functional and significant finite size effects. We use hysteresis in a number of measurable quantities to demonstrate a first-order transition up to a critical point, above which molecular and atomic liquids are indistinguishable. At higher temperature beyond the critical point, $H_2$-dissociation becomes a smooth cross-over in the supercritical region that can be modelled by a pseudo-transition, where the $H_2 \rightarrow 2H$ transformation is localized and does not cause a density discontinuity at metallization. Thermodynamic anomalies and counter-intuitive transport behavior of protons are also discovered even far beyond the critical point, making this dissociative transition highly relevant to the interior dynamics of Jovian planets. Below the critical point, simulation also reveals a dynamic $H_2 \leftrightarrow 2H$ chemical equilibrium with rapid interconversion, showing that $H_2$ and $H$ are miscible. The predicted critical temperature lies well below the ionization temperature. Our calculations unequivocally demonstrate that there are three distinct regimes in the liquid-liquid transition of warm dense hydrogen: A first order thermodynamic transition with density discontinuity and metallization in the sub-critical region, a pseudo-transition cross-over in the super-critical region with metallization without density discontinuity, and finally a plasma transition characterized by ionization process at very high temperatures. This feature and the induced anomalies originate in the dissociative transition nature that has a negative slope in the phase boundary, which is not unique to hydrogen, but a general characteristic shared by most dense molecular liquid. The revealed multifaceted nature of this dissociative transition could have an impact on the modeling of gas planets, as well as for the design of H-rich compounds.

# I. INTRODUCTION

Hydrogen is a simple element, but exhibits complex behavior at high pressures. Rich physics and chemistry have been discovered, and are still being predicted in both pure hydrogen [1][2][3] and hydrogen-rich compounds [4][5][6][7]. Dense solid hydrogen shows an unexpectedly complicated phase diagram [8][9][10][11][12][13][14][15] with an anomalous melting curve maximum and minimum [3][16][17][18][19][20][21], embodying solid states based around free rotating molecules (Phase I), broken symmetry due to quadrupole interactions (Phase II), packing of weakly bonded molecules (Phase III), and proposed "mixed" state (phases IV and V). The latter two phases have alternating layers of rotating molecules similar to Phase I, and weak molecules akin to Phase III [1][13][14][15]. Other phases under extreme compression include the recent claimed (and still controversial) molecular conductor or atomic metal [12][22][23], the predicted mobile solid state [2][21], and superconducting superfluid quantum liquid [24][25][26]. This wide range of behavior highlights the significance of dense hydrogen as an *archetype* of a many-body quantum system [23][24][27].

At sufficiently high pressure, liquid hydrogen becomes metallic. This is associated with the electronic transition from molecular $H_2$ to atomic H [28][29][30]. Historically, $H_2$ dissociation (*i.e.*, $H_2 \to 2H$) at high pressure was first proposed as a process that coincides with the *ionization* in which electrons leave the $H_2$ successively, namely, $H_2 \to H_2^+ + e \to 2H^+ + 2e$ [31]. The resultant state would be a plasma; therefore, the corresponding transition is termed a plasma phase transition (PPT) [32][33]. Recent high-pressure investigations, however, suggested that the molecular dissociation temperature should be related to *covalent bond-breaking* energies, rather than full ionization [27][34][35][36][37][38]. All experimental and simulation evidence indicates this liquid-liquid transition (LLT) and metallization can occur at temperatures well below the *ionization energy*. In other words, dissociation of liquid hydrogen at low temperature is intrinsically a unique phase transition *different* from the PPT.

The nature of this low temperature transition has been under debate for a long time [39][40][41][42][43][47]. Recent calculations based on density functional theory (DFT) and quantum Monte Carlo (QMC) suggested that it should be a *first-order* transition terminated at a *critical point* (CP), and be coincident with the metallization at low temperatures [35][44][45][46]. However, like ionization, a localized and uncorrelated thermally-activated process of bond-breaking should not yield a first order transition that involving a *collective* change; it must couple to other variables to induce the required large-scale correlations. Specifically, quantitative disagreement about the transition line remains. The critical temperature early estimated using DFT (via kinks in EOS) gave $T_c \approx 2000$ K [44][45]. A recent CEIMC estimation is between 1000~1500 K [35]. By checking the variation of proton-proton radial distribution functions (RDF) calculated with DFT, Norman *et al.* claimed an unusually high $T_c \geq 4000$ K [48]. A recent but not-well converged DFT simulation also reported a similar $T_c$ [49]. The latest VMC-MD estimation of $T_c$ was between 3000~6000 K, also identified via small kinks in EOS [46].

Another open question is whether dissociation involves $H_3^+$ cations or not [46][48][50]. This is important not only because they were used as a diagnostic to determine the transformation $T_c$ [46][48], but also because in design of H-rich superconductors it is a prerequisite to form large H-clusters or clathrate structures [4][5][6] that can be viewed as an *intermediate step* in hydrogen dissociation where the electron-phonon coupling being maximized.

Finally, the atomic and molecular H miscibility and transport properties near or below the critical point are still unknown, and require larger supercells than typically used to correctly describe liquid structure [59]. They are of great significance for modeling the convective flow crossing the $H_2$/H layer in giant gas planets [31][34]. So despite all these previous works, a central question remains whether all relevant physical quantities are discontinuous at a *single*, *first-order*, LLT in warm dense liquid hydrogen?

By using well-constrained and converged first-principles simulations, we addressed these important issues. The pressure of the LLT turns out to be extremely sensitive to the choice of exchange-correlation functional, while the critical temperature $T_c$ of LLT is better characterized and found to lie between 1000 and 1500 K. This supports the latest NIF experimental assessment [47] and the CEIMC result [35], but contradicts the previous DFT [48] and VMC-MD [46] assertion. In addition, $H_3$ clusters can be frequently identified by proximity of three atoms, but the lifetime is shown to be too short to produce any spectroscopic signature or for $H_3$ to be sensibly regarded as a chemical species.

For a first-order transition with a discontinuous density and electric conductivity at the LLT, one also expects a distinct two-phase coexistence interface. Nonetheless, we find that such phase separation is *impossible* because of the rapid $H_2$-2H interconversion. More importantly for planetary dynamics, we demonstrate a counter-intuitive increase in the proton self-diffusivity with pressure.

The paper is organized as follows. In Sec. II we present the methodology and computational details. The main results and discussion are given in Sec. III, which covers the topics of miscibility of $H_2$/H, a careful estimate of the critical temperature at ~1250 K at the DFT level, the low probability of $H_3$ cluster and their short lifetime, anomalous thermodynamics and proton transportation in the vicinity of the dissociative transition, a pseudo-transition model for this transition beyond the critical point (*i.e.*, in the super-critical region), and the three distinct LLT regimes. Finally, in Sec. IV, further discussion and potential impact to the interior dynamics of gas planets are presented, with Sec. V provides a summary of the main results.

## II. METHODOLOGY AND COMPUTATIONAL DETAILS

The first-principles calculations were carried out using DFT and a projector augmented-wave pseudopotential for the ion-electron interaction, and with two different exchange-correlation functionals for the electron subsystem: the generalized gradient approximation of PBE, and the van der Waals functional of vdW-DF (specifically, revPBE-vdW) [51][52], as implemented in VASP, to constrain the results. It is well recognized that PBE is deficient in describing $H_2$ metallization pressure [53][54]. But it can be expected that the true physics in dense hydrogen near dissociation is bracketed by PBE and vdW-DF, with the former underestimating the dissociation pressure whereas the latter over-stabilizes the $H_2$ molecule [55][56], as both the recent accurate CEIMC calculations [35] and dynamic compression experiments [27] suggested. At a fixed density near the dissociation, it was estimated that PBE (vdW-DF) underestimated (overestimated) the pressure by about 10~20 GPa in hydrogen. These two functionals are therefore employed simultaneously in this work to get a reliable assessment of the unknown systematic error in DFT.

In *ab initio* molecular dynamics (AIMD) simulations, supercells containing up to 3456 hydrogen atoms with periodic boundary conditions were employed. The radial distribution function (RDF) shows four well defined molecular shells which are captured in the 500-atom supercells which form the basis of our work, but would be destroyed by periodic boundary condition artifacts in simulations with smaller system sizes. The Baldereschi mean point is utilized for *k*-point sampling, which has been carefully checked for liquid hydrogen and gives an accuracy equivalent to a 4×4×4 mesh [39][45]. This setting is necessary to eliminate the possible spurious structures in MD generated by single Γ-point sampling [57][58]. By contrast, Ref. [49] employed a much smaller system with 64 atoms and a 3×3×3 *k*-point mesh, which is obviously not well converged. The energy cutoff for the plane wave basis set is 700 eV. The canonical ensemble (*NVT*) is used, with a timestep of 0.5 *fs*. The temperature is controlled by Nose-Hoover thermostat, and the conditions of thermodynamic equilibration and ergodicity are carefully checked. We find the liquids equilibrate much faster than solids, so after equilibrating for 1 *ps*, further AIMD simulations are then carried out for 1.5 *ps* to gather ensemble-averaged statistics. This enabled us to thoroughly explore finer P-T space, up to 500 GPa and from 500 to 3000 K. To check the sampling quality, some longer to 6 *ps* simulations were also done: these give smaller statistical fluctuations, but no different behaviors. The calculated PBE dissociation curve is in good agreement with previous estimates [27][44][45], which serves as a direct validation of our method for the following calculations.

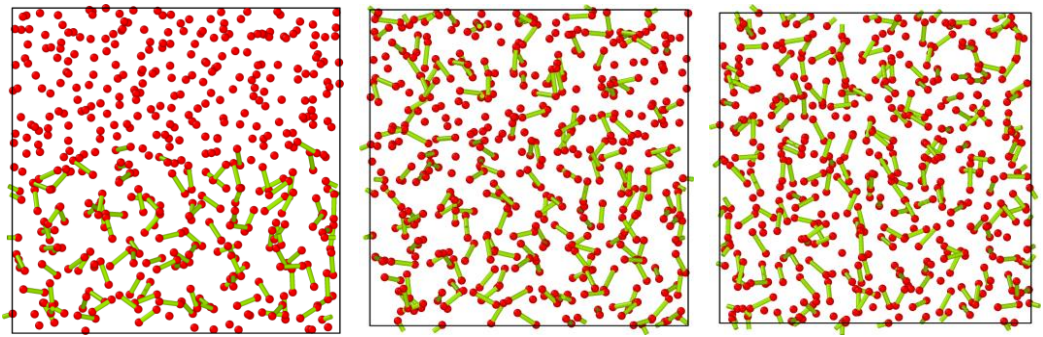

FIG. 1. (Color online) Mixing of atomic H and molecular $H_2$ within the dissociation region around 125 GPa and 1500 K, calculated by AIMD with PBE functional. Left panel: initial two-phase coexistence with a clear interface; middle panel: after 5 *fs* equilibrating new bonds have formed in the upper region and bonds have broken in the lower region; right panel: after 950 *fs*

equilibrating. The bond length cutoff criterion for drawing green lines is set as 0.9 Å.

## III. RESULTS AND DISCUSSION

### A. $H_2$/H interface and miscibility

A discontinuous first-order LLT [35][44][45] implies that the material is expected to have a two-phase coexistence. Nonetheless, it is worth pointing out that two-phase coexistence does not necessarily guarantee a distinguishable two-phase *interface* in the real space (*i.e.*, the occurring of phase separation). The latter appears only when the order parameters are quantities well-defined in real space, and the new phase *nucleates and grows* from a single (or just a few) embryo. Otherwise, if there are infinite embryos, the transition might manifest as being homogeneous rather than heterogeneous. The LLT in hydrogen is further complicated because $H_2$ and H are distinct chemical species, and the reaction between these must be in chemical equilibrium. If this reaction occurs much faster than phase separation, it will appear as miscibility.

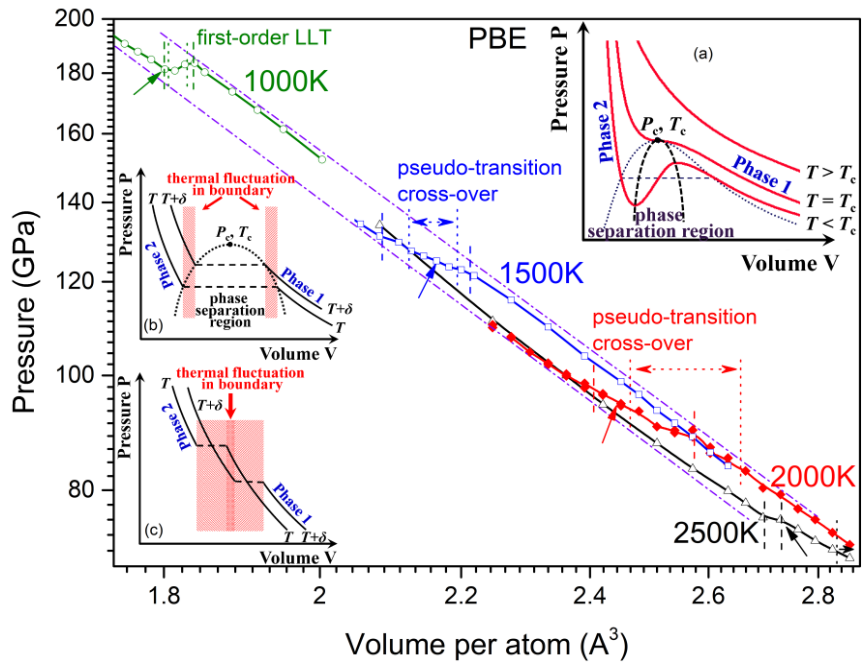

FIG. 2. (Color online) Isothermal pressure-volume curves of warm dense liquid hydrogen

across the dissociation region calculated with PBE functional. A first-order LLT with distinct hysteresis is obtained at only low temperatures. The dash-dotted long lines are guides to the eye only. Inset: (a) typical isotherms in the vicinity of the critical point for a normal first-order transition, below which there is a well-defined phase separation region that gives rise to a two-phase coexistence interface; (b) variation in the phase boundary driven by thermal fluctuations of case (a); (c) schematic of isotherms for $H_2$ dissociation, in which the boundary variation induced by thermal fluctuations could eliminate the prohibited region completely (*i.e.*, overlapping of the shaded regions).

To investigate the $H_2$/H miscibility, we carried out AIMD simulations in the atomic-molecular hydrogen coexistence region, using two-phase coexistence as the initial condition. This is a standard method to determine the first-order transition boundary such as melting [60][61]: the thermodynamically stable phase grows at the expense at the metastable one. As shown in Fig. 1, we find that the $H_2 \leftrightarrow 2H$ reaction is very fast, and the initial $H_2$/H interface disappears at the femto-second scale, far more quickly than the boundary could move even at sound speed. There is no growth, movement, or evolution of the interface boundary in AIMD simulations. Instead, it is the formation of atomic H *inside* the $H_2$ domain and vice versa that causes this LLT [59]. This *reversible* chemical reaction proceeds much faster than the nucleation and growth process could. It also suggests that under this condition the $H_2$ dissociation is mainly a *local* process in which the breaking or forming of individual $H_2$ dimers is insensitive to the local atomic environmental details. The same phenomena are also observed at 1000 K on the dissociation boundary, well below the previously estimated critical temperature [59].

This behavior is quite counter-intuitive for a typical first-order transition, but can be understood in terms of the unique thermodynamics of dense liquid hydrogen. Usually, a first-order phase transformation implies the existence of a density region, in which the total free energy is minimized by phase separation via binodal decomposition, if the density difference is a proper order parameter such as in the case of liquid-vapor transition (Fig. 2(a)). It establishes an interface when the phase boundary is robust against thermal disturbance (Fig. 2(b)). Nonetheless, the snapshots of Fig. 1 and the calculated isotherms in the main panel of Fig. 2 show that the LLT of hydrogen clearly *does not* belong to this case. The isotherms belong to a type of Fig. 2(c) rather than Fig. 2(b). The real space boundary of this type transition (*i.e.*, a two-phase interface) is *volatile* if subjected to thermal perturbations: it does not favor a phase separation in the pressure-density space, due to an intrinsic nature originated in the negative slope of the phase boundary on the P-T space.

This unusual behavior can be understood further by regarding the molecular and atomic hydrogen as two different chemical species (*i.e.*, viewing the transition as a chemical reaction). In this sense, the dissociative transition more or less relates to the concept of "non-congruent" phase transition [62]. In the conventional phase-separation region, they would have the same free energy and can interconvert without any free-energy penalty. However, the molecular phase can lower its free energy still further through the increased entropy of mixing of H and $H_2$ after partial dissociation into atoms, and vice-versa in the atomic phase. The equilibrium state is the same in both cases, *i.e.* the miscibility gap is wiped out, as the hatched areas in Fig. 2(c) shown. In a macroscopic picture for this kind of *abnormal* first-order LLT, the interfacial $H_2$/H free energy is negative, so the two-phase interface is volatile and forming additional interfaces is favored. In terms of nucleation and growth, it means the critical nucleus is infinitely small, so $H_2$ dissociation *proceeds* spontaneously and *homogenously with infinite micro-embryos, and does not sustain a distinct and stable two-phase coexistence interface*. This interesting scenario is further corroborated by the partial negative correlations in the bond-length of nearest neighboring $H_2$ dimer along the dissociation [59], which also support $H_2$/H miscibility even below the critical point (*i.e.*, in the sub-critical region). It should be stressed that our calculations strongly support the picture that as one-component system hydrogen realizes as "molecular" or "atomic" state depending on the P-T conditions, and in the transition region, hydrogen "molecules" are transient bound states or short-lived correlations, as their lifetime shown in [59] reveals.

The abnormality in $H_2$ dissociation also affects the thermodynamics in the vicinity of the dissociation region even *far beyond* the critical point where a conventional supercritical fluid should already become normal. For the isotherm calculations shown in Fig. 2, we find both thermal expansion coefficient and pressure coefficient display a pronounced abnormal dip, reaching a negative value of about $-6 \times 10^{-5} K^{-1}$ and $-1.2 \times 10^{-4} K^{-1}$ at 2000 K, respectively, as shown in Fig. 3. By contrast, the compressibility has a peak in the dissociation region, which can be understood as being due to the $H_2$/H reaction providing an extra mechanism by which the liquid can densify. All of these are due to the particular variation of EOS across this dissociative (or pseudo-transition) region. As shown in Fig. 2, the isotherm of 2000 K has two intersections with the 1500 K curve (other isotherms are similar). This is a common feature for all transitions or cross-over that induce a *softening* in the compression curve but at the same time has a *negative* slope in the phase boundary on the P-T space. It should be noted, however, that this is *not* a common feature for all first-order transition that terminates at a critical point, as well as the corresponding super-critical behavior.

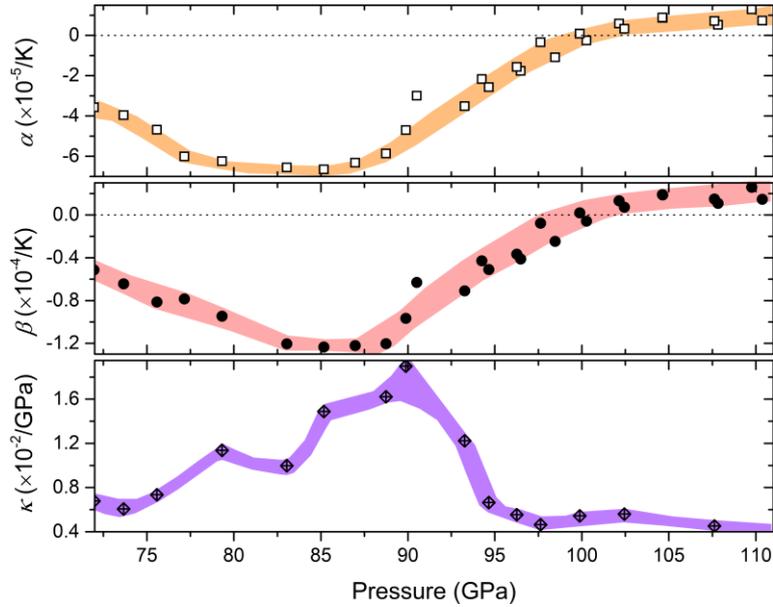

FIG. 3. (Color online) Estimated thermal expansivity $\alpha = \frac{1}{V}\left(\frac{\Delta V}{\Delta T}\right)_P$, pressure coefficient $\beta = \frac{1}{P}\left(\frac{\Delta P}{\Delta T}\right)_V$, and compressibility $\kappa = \frac{-1}{V}\left(\frac{\Delta V}{\Delta P}\right)_T$ at 2000 K by using the PBE isothermal data, respectively. Note the striking *negative* thermal expansivity and pressure coefficient, as well as the anomalously *peaked* compressibility across the dissociation region, indicative of a continuous pseudo-transition crossover above the critical point rather than a thermodynamic phase transition. Solid lines are guides to the eye only.

This intriguing behavior is similar to a *pseudo-transition* in non-stoichiometric compounds, which is inherently continuous, whereas a rapid cross-over of the free energy leads to a sharp abnormal variation in the thermodynamics [63][64]. In fact, dimer dissociation at the dilute limit has the same mechanism of pseudo-transition [59]. On the other hand, our results showed that these anomalies occur in a broad thermodynamic region both below and beyond the critical point of dense liquid hydrogen. It is a general feature for dissociative transition of dense molecular liquid which has a negative slope of phase boundary, and is highly relevant to the interior condition of Jupiter and Saturn, and so could have a profound impact on the magneto hydrodynamics modeling of convective flows in these planets.

## B. Critical point of the LLT

As demonstrated in Fig. 2, the 1000 K isotherm displays a sharp jump and hysteresis at the LLT (a similar result also holds for vdW-DF [59]). This strong signature of a first-order transition unequivocally proves that $T_c \geq 1000 \text{ K}$. At temperatures higher than 1500 K, however, the hysteresis vanishes, and the identification of the nature of the transition requires examining higher derivatives of the free energy, such as heat capacity. Previous claims for the first-order transition (and thus to determine $T_c$) were based only on a sharp change (or kink) in the isotherms via visual judgement [27][35][44][45][46]. Unfortunately, identifying a "volume collapse" or kink on a P-V curve based on measurements or calculations that carried out at discrete volumes is insufficient to validate it as a first-order transition or not [63][64]. For example, it is hard to tell whether the erratic variation in the 2500 K isotherm (see the arrow in Fig. 2) is a kink or not. The conclusion depends sensitively on the interval between the discrete data, as well as on the numerical accuracy of pressure measurement. This ambiguity could be one of the reasons for the scattering in the reported $T_c$ estimated using EOS kinks. If we apply the same criterion as used previously, our AIMD data would also give an unphysical $T_c$ higher than 2500 K.

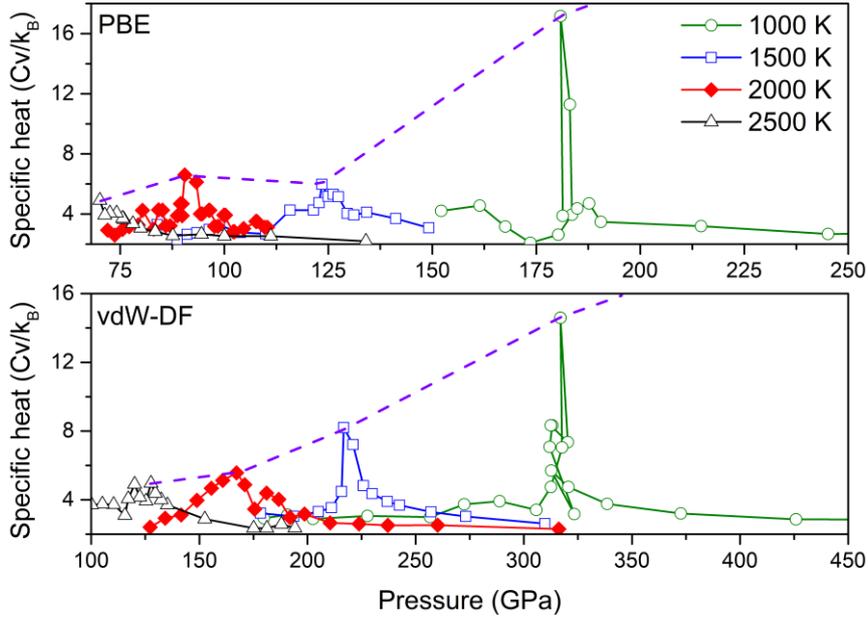

FIG. 4. (Color online) Specific heat extracted from thermal fluctuations in AIMD simulations at given temperatures calculated with PBE and vdW-DF, respectively. Dashed lines that connect peaks are guides to the eye only.

Another method that used previously to identify the LLT and to locate $T_c$, is the relative variation of RDF [46][48]. The drawback of this approach is that it implicitly assumes the $H_3$ cluster as a well-defined stable species. As will be seen below, this is not the case in dissociating hydrogen where the RDF evolves continuously with pressure. Therefore, the method to trace the relative variation of the RDF difference between its first peak and first valley becomes an arbitrary criterion, since one can instead choose the second peak and third valley, or any point along the radial distance, as the reference points for the transition. A convincing signature of a first-order transition is *hysteresis* which was not considered in Refs. [46][48]. Such hysteresis—two densities observed at the same pressure—can be seen in the 1000 K isotherm in Fig. 2.

In order to locate $T_c$ more reliably, we also employ a different method. It is well-known that when approaching a phase transition, thermodynamic fluctuations become large, even being divergent in the case of a first order transition and closing to the critical point. In a practical AIMD simulation, due to the finite size of the employed supercell, one cannot get a true divergence. However, the fluctuation magnitude could become exceptionally large, so that its variation is sensitive enough to pin down $T_c$ precisely.

According to the fluctuation–dissipation theorem, fluctuations in energy give the specific heat. Such calculated specific heat is plotted in Fig. 4. It shows that at 1000 K the transition is sharp and narrow, being consistent with a first order transition that approaching the CP. However, it becomes broad and smooth when above 1500 K, illustrating both the width of the cross-over region and the position of the Widom line in the supercritical region. The divergence disappears somewhere between 1000 and 1500 K for both PBE and vdW-DF (with the former being more distinct whereas the latter being more progressive). This provides unequivocal evidence that a critical point exists for the low temperature LLT, and $T_c$ should be in this range. This value, ~1250 K, is in good agreement with recent CEIMC assessment [35] and the latest NIF dynamic experiment that observed a sharp transition when below 1080 K but did not resolve a reflectivity jump when beyond 1450 K [47]. As mentioned above, the actual dissociation should be bracketed by the results of PBE and vdW-DF, Fig. 4 therefore refutes any theoretical $T_c$ higher than 1500 K [46][48]. It is worth mentioning that our data are also in good agreement with previous PBE results of [45], which revealed a density jump below 1500 K that is driven by an abrupt dissociation with a jump in electrical conductivity, showing the characteristic of a nonmetal-to-metal transition along with dissociation of $H_2$ molecules. This conclusion marks a *consensus* on $T_c$ between DFT and QMC, as well as between the theory and dynamic compression experiment.

It should be mentioned that because the critical point is close to the melting curve, one might worry the possible interference of the results from the metastable crystalline-like structures. We carefully checked the simulations of 1000 K, and did not find any signature of crystalline-like patterns. It also should be pointed out that using the similar DFT setting, we obtained a melting curve in good agreement with other simulations and experimental data [3][20]. Our estimated critical temperature here is ~600 K higher than the melting curve.

### C. Possibility of $H_3^+$ cluster

In addition to $H_2$ dimers, larger H-clusters have frequently been predicted as important components in compressed H-rich compounds [5][6][7] and ultra-dense solid hydrogen [3]. One of the most common is the $H_3$ unit. Ref. [48] employed a geometric feature in the RDF corresponding to $H_3^+$ as a criterion to locate $T_c$. This treatment implicitly assumes $H_3^+$ is more important than any other clusters, and should be a stable chemical species (otherwise one cannot define a new "phase" if the characteristic feature is short-lived and all related thermodynamics thus must be continuous). Statistical analysis of the CEIMC data suggested that $H_3$ might not be as stable as previously assumed [50]. Its lifetime, however, has never been explicitly calculated, nor its valence state.

Using AIMD simulations, we obtained the lifetime of individual $H_3$ clusters, as well as their concentration with temperature and pressure across the dissociation region. The result shows that the lifetime of $H_3$ unit is actually very short (at a level of several *fs*) [59]. They are too unstable to be regarded as a chemical species. Objects identified as $H_3$ based on distance criteria [48] are typically transient encounters between $H_2$ and H, such as scattering, or intermediate states of reactions where one proton displaces another in the dimer. This supports the CEIMC assessment [50]. Furthermore, our Bader charge analysis shows that $H_3$ is not positively charged with a fixed valence state of +1 as usually expected. Instead, it is averagely *neutral*, with a charge state fluctuating frequently between ±δ (with $\delta \ll 1$), depending on its rapidly evolving geometry [59]. These observations indicate that the assignment of protons to the big but short-lived clusters is arbitrary, and might dismiss any evaluation of $T_c$ by referenced to $H_3^+$ ions as Ref. [48]. These transient clusters do present and manifest in RDF, which however is not enough to unambiguously identify a *thermodynamic phase transition* (*i.e.*, a sharp change in this short-lived quantity cannot generally generate a macroscopic discontinuity in the thermodynamic limit).

Besides $H_3$, we also observed other larger clusters. All of them have very short lifetime, and with strong fluctuations in their charge state or polarization [59]. This illustrates that the dissociation is not via a two-step mechanism as proposed in Ref. [48]. Rather, it comprises *multiple* transient and micro consecutive steps of $H_2 \rightarrow H_3 \rightarrow H_n \rightarrow \cdots \rightarrow H$. It also suggests that the complex H-clusters observed in H-rich compounds should originate from a mechanism that heavy elements acting as electron donors or acceptors to balance the charge distribution, so that stabilizes certain type of the H-clusters within the multiple-steps of dissociation process. It should be noted that the charge fluctuation or sloshing in H-clusters might lead to novel optical properties. For example, it will active and enhance the otherwise prohibited infrared/Raman modes, and give rise to a strong *noisy optical background* in the dissociating layer of liquid H in Jovian planets.

### D. Proton transportation

The sharp first-order LLT in dense hydrogen is associated with discontinuities in transport properties. It was shown that electric conductivity [39] and optical reflectivity [47] jump there, presumably due to metallization. It is thus natural to *expect that the transport of protons should also be discontinuous in the vicinity of*

*dissociation.*

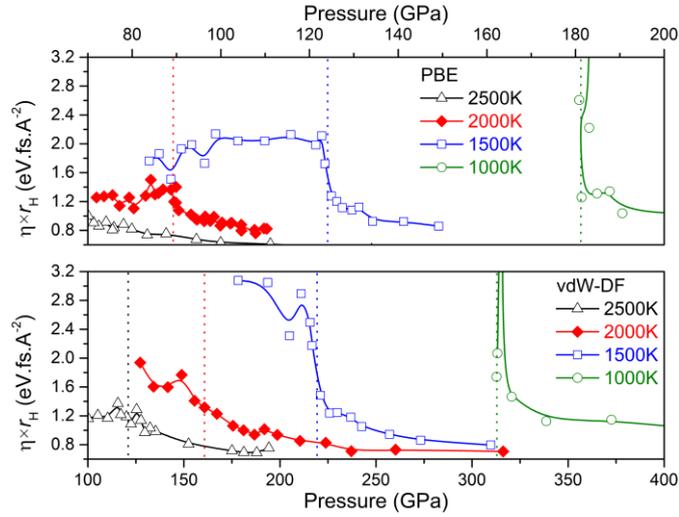

FIG. 5. (Color online) Variation of the viscosity $\eta$ (with respect to atomic hydrogen radius $r_H$) of warm dense liquid hydrogen with pressure across the dissociation region calculated from diffusivity using the Stokes-Einstein relationship, using both PBE and vdW-DF, respectively. The vertical dotted lines indicate the position where half of $H_2$ have been dissociated.

The AIMD-calculated viscosity based on proton diffusivity using the Stokes-Einstein relationship is shown in Fig. 5. Its variation is quite atypical. With increased pressure, viscosity usually increases, accompanying a reduction in particle mobility. Nonetheless, here we observed a drastic reduction in viscosity with increased pressure when crossing the dissociation region along the isotherms (this effect is equivalent to an enhancement in proton mobility), which saturates to the atomic value after full dissociation [59]. This abnormal decrease in viscosity can be understood by recognizing that lighter, smaller, dissociated H-atoms migrate faster than $H_2$ molecules. The pressure where the shift in viscosity occurs depends on the functional, but if viscosity is plotted against fraction of stable $H_2$ dimers, the results are independent of functional.

At the same time, we did not observe any discontinuity or kink in the self-diffusivity of hydrogen at or near $\rho_{H_2} = 0.5$ [59]. This is also out of usual expectation, showing the proton mobility is insensitive to the sharp first-order LLT. Overall, our calculation reveals that the proton diffusivity depends more sensitively on the fraction of H rather than H$_2$, and the rapid increase in proton diffusivity when $\rho_{H_2} \to 0$ [59] can be understood via a percolation mechanism [65][66].

### E. Pseudo-transition model

Because of the importance of H$_2$-dissociation even far beyond $T_c$ as shown above [31][34], it is necessary to derive a thermodynamic model to describe this broad and smooth super-critical cross-over (including the accompanying anomalies), which is mainly governed by *local* energetics, rather than by collective correlations. It is surprising that the variation of the dissociation width with pressure and temperature qualitatively matches a scenario of *pseudo-transition model* (PTM) [63][64], which initially was proposed as a thermodynamic cross-over in non-stoichiometric compounds that is continuous *a priori*, but at low enough temperatures can generate sharp kinks in some physical properties, resembling a typical first-order transition [63].

In the case of H$_2$-dissociation, if we ignore all *local* atomic environment effects, the reaction $H_2 \rightleftharpoons 2H$ can be viewed at the zero-order approximation (*i.e.*, the dilute approximation) as an equilibration in a two-energy level system. The principles of statistical mechanics then give a dimer fraction of $\rho_{H_2} = \left(\exp\left(\frac{-\Delta E}{T}\right) + 1\right)^{-1}$, which is identical to the result of a PTM [59]. Here $\Delta E$ is the binding free energy of H$_2$ dimers that is a function of pressure and temperature in general, and can be approximated as $\Delta E(P,T) = a - bP - cT - dTP$. The temperature dependence mainly comes from the contribution of excess entropy before and after the dissociation (*i.e.*, of the 2H against $H_2$). The pressure dependence comes from bond weakening, from several eV at low pressure approaching zero at about the dissociation pressure.

This PTM captures the main characteristics of the dissociation, as shown in the inset of Fig. 6(a). Reasonable parameterization [59] gives a progressive cross-over above 1250 K, but converges to a sharp LLT when below 1000 K. This AIMD-calculated change of the dissociative transition nature is understandable, since at low temperature and high density the orientation correlation among $H_2$ dimers is strong, whereas it becomes much weaker at higher temperatures and lower densities [67][68]. The vanishing of hysteresis in the high temperature dissociation region also supports this argument. A schematic representation of this multifaceted scenario is provided in Fig. 6(c).

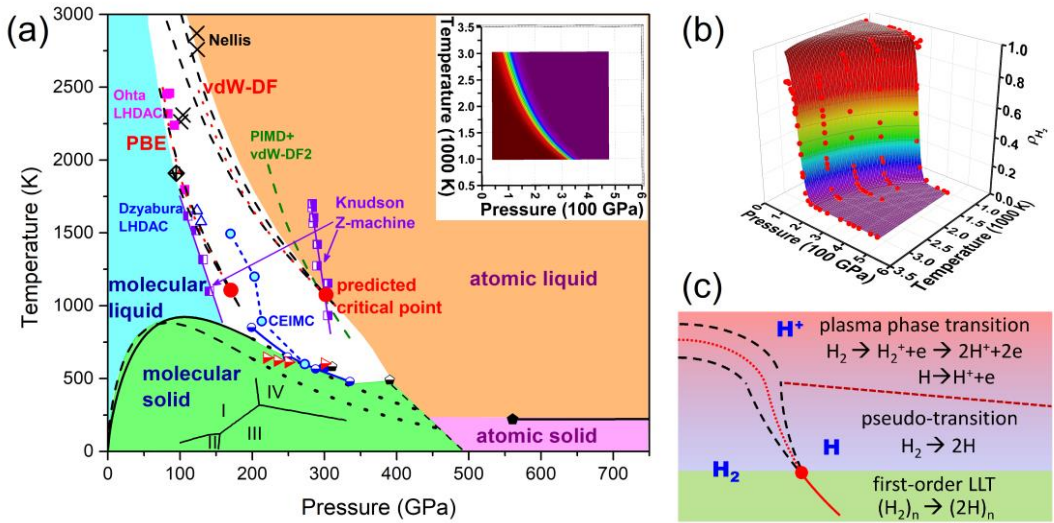

FIG. 6. (Color online) (a) Phase diagram of warm dense hydrogen around the dissociation region, with PBE and vdW-DF results bracket the true dissociative transition boundary. The bold black dashed lines represent the 3/7 lines of the dissociation cross-over, the red dotted lines indicate the extension of the first-order LLT boundaries (*i.e.*, the Widom line), with the estimated CP denoted by the red solid circle points, as predicted by PBE and vdW-DF, respectively. Inset

shows the dissociation region predicted by vdW-DF fitting to a PTM. The half-filled triangles are the melting points reported by Zha [18]. Note at high temperature limit, both PBE and vdW-DF converge towards the same dissociation curve. The MD-VMC data of Mazzola *et al.* [69] are not included here due to their poor quality of convergence. Other data: CEIMC—Pierleoni [35], PIMD+vdW-DF2—Morales [53], half-filled circles—Belonoshko [19], filled and half-filled pentagons—Geng [3][20], half-filled squares—Knudson [27], open triangles—Dzyabura [37], filled squares—Ohta [38], crosses—Nellis [70]. (b) $H_2$ dimer fraction predicted by a PTM compared to the *ab initio* values. (c) A schematic of the three regions of the liquid-H phase diagram, labelled by the dominant species $H_2$, H and $H^+$, and indicating the first-order LLT below $T_c$ and the crossover transitions (including pseudo-transition and PPT) elsewhere.

Figure 6(a) plots the phase diagram of warm dense hydrogen near dissociation. Relevant experimental data available so far are also shown. Zha's melting data [18] constrained the dissociation region from below, and are in good agreement with Geng [3][20] and Belonoshko's [19] theoretical estimation at about 300 GPa. Our dissociation line calculated with PBE is in remarkable agreement with the laser heating DAC results [36][37][38]. This is probably due to an error cancellation with too-weakly bonded PBE molecules compensating for the absent zero-point energy (ZPE) weakening; moreover, the DAC data are still under debate [47]. The CEIMC estimate [35], which includes both ZPE and QMC exchange-correlation effects at the expense of describing the two "liquids" with a total of 27 molecules, lies mid-way between our PBE and vdW-DF results, and in good agreement with the laser-shock measurement of the reflectivity [47]. The latter data are not shown here for the sake of clarity. In order to show the width of cross-over region, the AIMD-calculated 3/7 lines (that corresponding to an $H_2$ dimer fraction of 30% and 70%, respectively) are also plotted. These lines, together with the PTM results, narrows the uncertainty range of $T_c$ for the first-order LLT further down to between 1250 K and 1000 K, far below the previous DFT estimate of ~2000 K (the crossed rhombus point in Fig. 6) [27][44][45]. This new DFT estimate, however, is in good agreement with the latest experimental [47] and CEIMC [35] data. Above this point, the dissociative transition has a *finite width* on the P-T plane, and the boundary cannot be characterized by a single line any more.

## IV. FURTHER DISCUSSION

The assignment of the low-temperature dissociation and metallization transition in dense liquid hydrogen as a plasma phase transition has its historical reasons, but our work shows this to be untenable. The electron localization functions (Fig. 7) demonstrate that even in the metallic state the electrons are strongly associated with the ions, whereas the plasma transition should denote the ionization process of $H_2 \rightarrow H_2^+ + e \rightarrow 2H^+ + 2e$. The nature of this latter ultra-high temperature transition, however, is completely different from what occurs at just above the critical point. As shown in Fig. 7, the electrons are still localized around the atomic nucleus or covalent bond centers at these conditions. The dissociation and metallization processes in both sub-critical and super-critical regions are via orbital overlapping and the subsequent (partial) electron delocalization, rather than ionization. This observation is corroborated by charge analysis, where some atomic hydrogen are even negatively charged [59], strong evidence that it is not an ionization process. For this reason, we suggest to reserve the name PPT for the transition at ultra-high temperature that is obtained by kinetic ionization process and extends to ultra-high solar-pressures; between PPT and the critical point (*i.e.*, the super-critical region), the transition is a continuous cross-over, characterized by orbital overlapping and electron delocalization, but with very weak intermolecular correlations, which we would like to call it *pseudo-transition* to emphasize its unique dimer-dissociative characteristics that are not shared by normal super-critical cross-over; below the critical point, the transition is also driven by electron wavefunction overlap and delocalization, but now with strong intermolecular correlations and the resultant discontinuities in density and other physical quantities, and this regime in the sub-critical region is a *first-order LLT*. These *three* distinct regimes of the dissociative liquid-liquid transition are schematically shown in Fig. 6(c). It provides a comprehensive scenario for the temperature-pressure-driven metallization and ionization in warm dense liquid hydrogen.

It should be pointed out that the dimer fraction in $H_2$ dissociation is the proper order parameter for the transition, which correctly reproduces the prohibited region in the order parameter space as required by Landau theory of phase transition [59]. Most previous studies on this topic tried to understand the transition in the density space. Unfortunately, the density difference between the atomic and molecular liquid phases is *not* a proper order parameter for this dissociative transition. This, together with the pseudo-transition nature of the dissociation which also could lead to a continuous but "sharp" variation in physical quantities, contributed to the controversial nature and scattered CP location of this transition as reported in literature. The finite size effect complicates this further [59].

There is a general consensus on the coincidence of metallization and dissociation in dense liquid hydrogen. By using dimer fraction as an order parameter, we showed in Fig. 6(b) that beyond the thermodynamic critical point of ~1250 K, the dissociative transition has a *finite* width on the P-T plane, so *cannot* be a first-order transition. However, one might argue that the first-order transition could be different from the dissociative transition, thus can coexist with the latter simultaneously. Such a hypothesis that there is another first-order transition existing *within* the broad dissociative region beyond the critical point, as shown by the dotted red line in Fig. 6(c), is intriguing. The band gap appears to drop continuously to zero which does not imply any first order transition. The region defined by an electron density isosurface undergoes a percolation transition, but at different densities depending on its chosen value. We did not find any other order parameter with discontinuity in our AIMD simulations above the critical point. This can be understood by recognizing that the collective motion in any first-order transition must come from some correlations. As we showed above, intermolecular correlation is very weak in liquid hydrogen [59]. The main contribution is the angular orientation arising from compression effect. At low pressure, inter-molecule distance becomes large enough so that it loses all orientation correlations. This can be seen clearly in Fig. 6, in which the CP of PBE is at such a low pressure whose low temperature region is occupied by phase I that does not have any dimer orientation correlations. That is to say, in the liquid state at high temperature of the same density or below, it is unlikely that there still have inter-particle correlations (other than the dissociation itself) to cause a first-order transition.

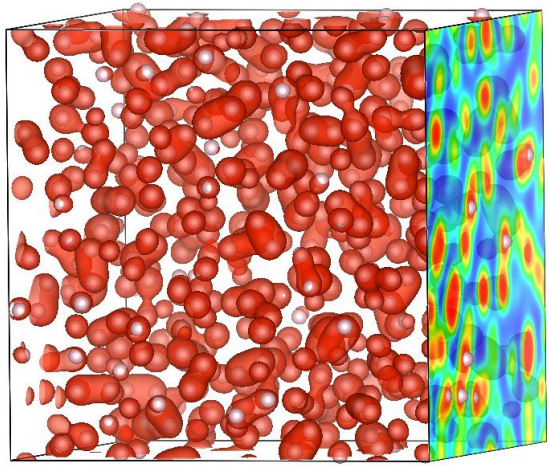

FIG. 7. (Color online) Electron localization function isosurface (with ELF=0.92, in red) for liquid $H_2$ within the dissociation region at 138.5 GPa and 1500 K (with an atomic volume of 1.9512 Å$^3$/H) calculated by PBE. An ELF plane is also included [ELF going from 0 (blue) to 1 (red)].

The results we revealed above might have an impact on the modeling of the convective flow across the $H_2$/H layer and the dynamo of cold giant gas planets [34]. The low $T_c$ of both our DFT and CEIMC implies the irrelevance of the first-order LLT to the interior of Jupiter and Saturn. Even if this transition presents in a very cold gas planet, our calculation predicted that the $H_2$/H interface is volatile, and $H_2$ and H are *miscible*. Its direct influence on the convective flow is thus much less significant than previously estimated. On the other hand, our results also suggested that strong thermodynamic anomalies (*e.g.*, negative thermal expansion) extend far above $T_c$. This P-T condition is directly relevant to the interior of Jupiter and Saturn. For example, along Jupiter's adiabat, DFT calculations revealed anomalous peak in heat capacity and dip in thermal expansivity that occur exactly in the dissociation region [71]. This is qualitatively in line with our results as presented in Figs. 3 and 4 even though the composition is different and the temperature is much higher and the pressure is lower. The hydrogen diffusivity along Jupiter's adiabat as reported in [71] also jumps when crossing the dissociation region, and the numerical value near the dissociation is comparable to our data. It confirms that anomalies associated with dissociation can extend very far beyond the critical point. This striking "wide-range" influence arises from the unique behavior of molecular dissociation (*i.e.*, the pseudo-transition mechanism), and cannot be accounted for by *normal* super-critical behavior. The observed thermodynamic anomalies, together with the predicted anomalous drop of viscosity across the dissociation, contribute to the thermal instability of convection cells and internal dynamics of cold giant gas planets. For example, the large-scale convection cell in gas planets could be cut by this anomalous layer into two parts, changing the convection flow cycle from a usual "O" shape into an "8" shape, and resulting in an advection layer in between [59].

# V. CONCLUSION

In summary, the complex nature of liquid-liquid transition in hydrogen was investigated with AIMD simulations. We find a first order thermodynamic transition line which terminates at a critical point. Above the critical point, the molecular-atomic transformation causes anomalies in the viscosity and thermal expansivity.

This broad and smooth super-critical cross-over region, and the accompanying anomalies, can be described by a pseudo-transition model. Going from low pressure to high pressure, compression enhances inter-particle interactions, weakens the covalent bonds and modifies the dimer binding energy. This lowers the corresponding dissociation temperature from ionization energy to much lower temperature. Compression also puts strong constraints on molecular rotations, subsequently enhances *local* angular orientation correlations, leading to large but transient H-clusters during dissociation. At low enough temperature, the dissociative transition eventually develops into a first-order LLT when below ~1250 K. Unlike typical first-order transitions, the $H_2$/H two-phase coexistence interface of this LLT is unstable, and the parent-daughter phases are *miscible*, which is a general feature for such a dissociative transition with a large negative slope of $dT/dP$ along the boundary. This density-driven LLT is comparable to the solid Phase I-III-metal transition, while the miscibility of H/$H_2$ species is reminiscent of the atomic-molecular solid Phase IV. Finally, other facets of the transition as discovered in this work, *e.g.*, the thermodynamic anomalies due to pseudo-transition and the counter-intuitive variation of the proton diffusivity and viscosity in the vicinity of dissociation, could have a significant impact to the modelling of the interior of cold Jovian planets.

**Acknowledgement.** This work was supported by the National Natural Science Foundation of China under Grant Nos. 11672274 and 11274281, the NSAF under Grant No. U1730248, the foundation of National Key Laboratory of Shock Wave and Detonation Physics of China under Grant Nos. 6142A03010101 and JCKYS2018212012, and the CAEP Research Projects CX2019002. We thank the UKCP Archer computing service at EPCC (EPSRC grant K01465X). G.J.A. and M.M. acknowledge support from the ERC fellowship "Hecate" and a Royal Society Wolfson fellowship. Part of the computation was performed using the supercomputer at the Center for Computational Materials Science of the Institute for Materials Research at Tohoku University, Japan. H.Y.G. appreciates Prof. Roald Hoffmann of Cornell University and R. J. Hemley of University of Illinois at Chicago for helpful discussions.

**Supplementary Material.** Results of dimer-dimer bond-length distribution (DDLD) calculations, comparison between PBE and vdW-DF results, the thermodynamic modelling of pseudo-transition and phase boundaries, calculated isotherms (equation of state, EOS), transient clusters and their lifetime and charge state, the angular distribution function of $H_3$ clusters, mixing of H and $H_2$ liquid, thermal fluctuation analysis, isotope effect, finite size effects, and the assessment of proton self-diffusivity and viscosity.

**Competing financial interests**

The authors declare no competing financial interests.

# Supplementary Material for "Thermodynamic anomalies and three distinct liquid-liquid transitions in warm dense liquid hydrogen"

Hua Y. Geng[1,2,*], Q. Wu[1], Miriam Marqués[3], and Graeme J. Ackland[3]

[1]*National Key Laboratory of Shock Wave and Detonation Physics, Institute of Fluid Physics, CAEP; P.O.Box 919-102 Mianyang, Sichuan, P. R. China, 621900*

[2]*Center for Applied Physics and Technology, HEDPS, and College of Engineering, Peking University, Beijing 100871, China*

[3]*CSEC, SUPA, School of Physics and Astronomy, The University of Edinburgh, Edinburgh EH9 3JZ, United Kingdom*

## I. Dimer-dimer length correlations (DDLC) of vdW-DF functional

In the condensed state under high pressure, there are several structural correlations between molecules: (*i*) dimer-dimer length correlations (DDLC), (*ii*) molecular orientation rearrangements, (*iii*) bond angles between triplets, etc. Among these, the DDLC is the most relevant quantity for the LLT since the essential aspect in $H_2$ dissociation is the variation of the dimer length: it is governed by the stretching vibrational mode. We examine the collective variation of bond length between dimers to see if there is any collective dissociation in molecular $H_2$. These involve comparing the bond lengths of neighboring molecules to determine if molecular stretching is correlated. We then build a 2D correlation map relating the length of a dimer to the length of its neighbor. We look for four types of dimer-dimer length correlation (DDLC), as shown schematically in Fig. S1(a~d). A strong nucleation and growth mechanism (NGM) requires the dimer-dimer length distribution (DDLD) being positively correlated [Fig. S1(c)]. This is easy to understand, since the transition is actually a reaction of $H_2 \rightleftharpoons H + H$ during which the characteristic change is the dimer length. It also means that a featureless DDLC [Fig. S1(a)] denotes a dissociation that is independent of the local atomic environment, whereas a negative DDLC [Fig. S1(b, d)] implies that nucleation and growth of the atomic H phase out of $H_2$ phase is not favored.

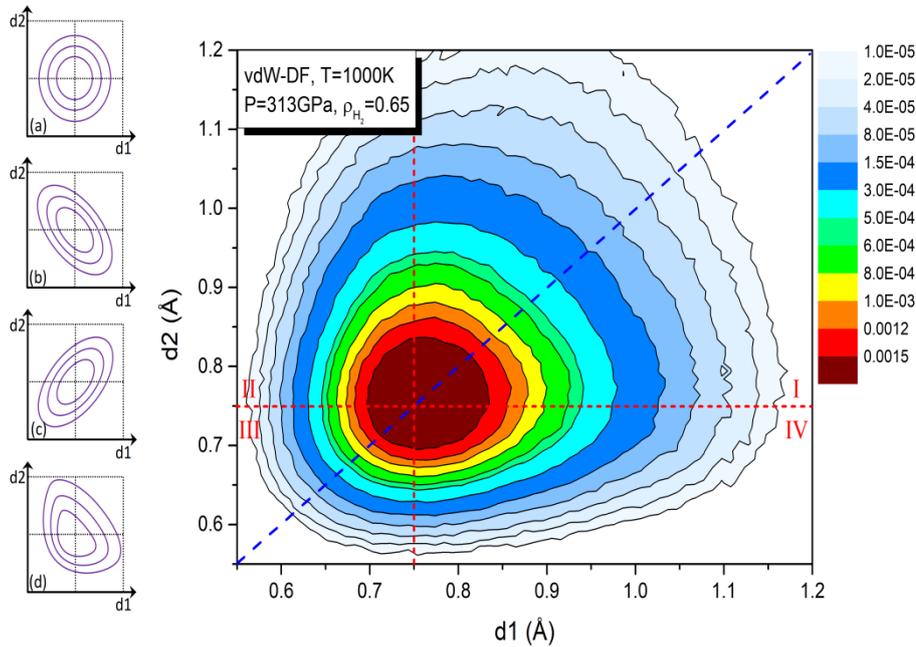

FIG. S1. Dimer-dimer length distribution (DDLD) of warm dense liquid hydrogen in the dissociation region for two nearest-neighboring independent $H_2$ dimers (here "dimer" means any pair of atoms remaining within 0.95 Å for more than 75 *fs*, after excluding any contribution from $H_3$ clusters). Left panels are schematics of typical DDLD for (a) no dimer-dimer length correlation (DDLC), (b) negative or anti-correlated DDLC, (c) positive DDLC, and (d) partially anti-correlated.

The main panel in Fig. S1 shows a DDLD of dense liquid hydrogen at 313 GPa and 1000 K calculated by vdW-DF (PBE results are similar). This P-T condition is on the LLT boundary and well *below* all previously estimated $T_c$, where the transition was determined to be first-order. It is interesting to note that, in contrast to usual expectation, this DDLD belongs to the type of Fig. S1(d), rather than the type (c). Or explicitly, the dimer length of nearest neighboring $H_2$ dimers is partially anti-correlated, *i.e.*, those nearly-dissociated dimers are mainly surrounded by shorter ones. Perceptible positive correlations are observable only when $H_3$ are also included, as shown in Fig. S2, indicating the dominance of symmetric stretching vibration in $H_3$ units.

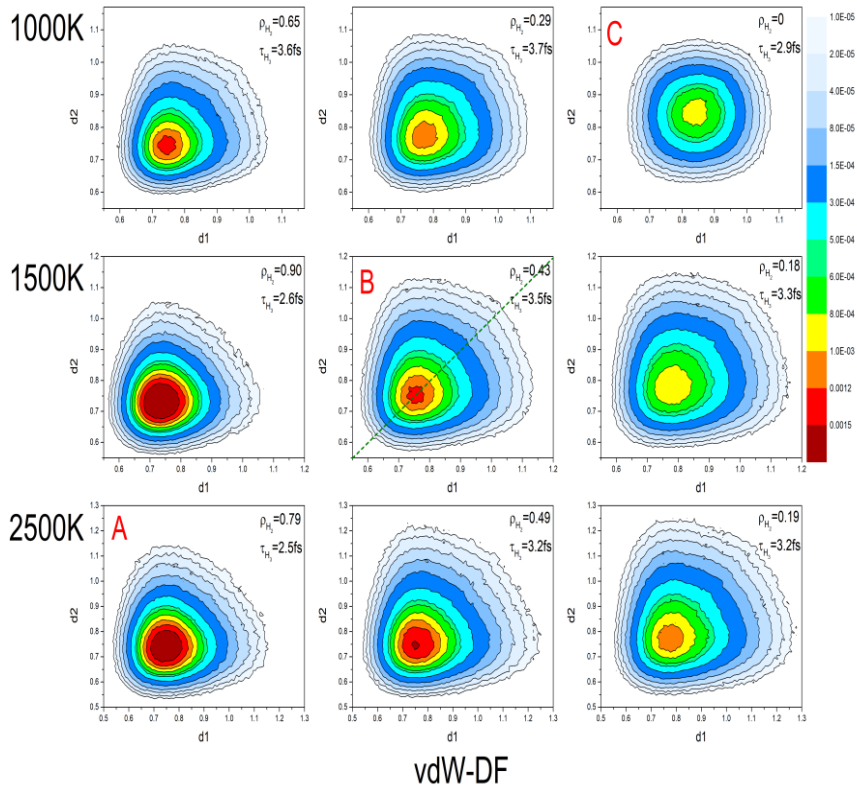

FIG. S2. Dimer-dimer length distribution (density of probability, including $H_3$ contribution) in warm dense liquid hydrogen around $H_2$ dissociation. The calculations were performed with vdW-DF. Note that though vdW-DF and PBE predict quite different dissociation pressures and temperatures, their dimer-dimer correlations are actually very similar. A negative correlation for the stretching modes is observed at the initial stage of dissociation (panel A): the nearest neighboring $H_2$ dimer tends to have a short bond length when the reference dimer is stretched to dissociating. Then a positive correlation develops (along the diagonal direction in the figure) when more than one third of $H_2$ molecules have been dissociated (panel B): the nearest neighboring $H_2$

dimer tends to have a long bond length when the reference one is stretched (note that this occurs *only* when they are in the same $H_3$ cluster). After all molecules are fully dissociated, there is no appreciable length correlation between neighboring dimer pairs any more (panel C).

The dimer-dimer length distributions (DDLD) as presented in Figs. S1 and S2 imply that for hydrogen dissociation at high temperatures an independent but elongated or dissociating "dimer" is surrounded mainly by shorter dimers. In other word, the atomic and molecular hydrogen are miscible. It is the opposite of what is needed for phase separation in the real space. It also indicates that the dissociation is mainly a *localized* process for each individual $H_2$ dimer, almost independent of its local environmental details, except for the *bulk* compression effects. $H_2$-dissociation thus can be viewed as a homogeneous thermal excitation process approximately (dilute limit approximation). The weak correlations among $H_2$ molecules at high-density and low-temperature that drive the first order LLT should originate in the angular orientation.

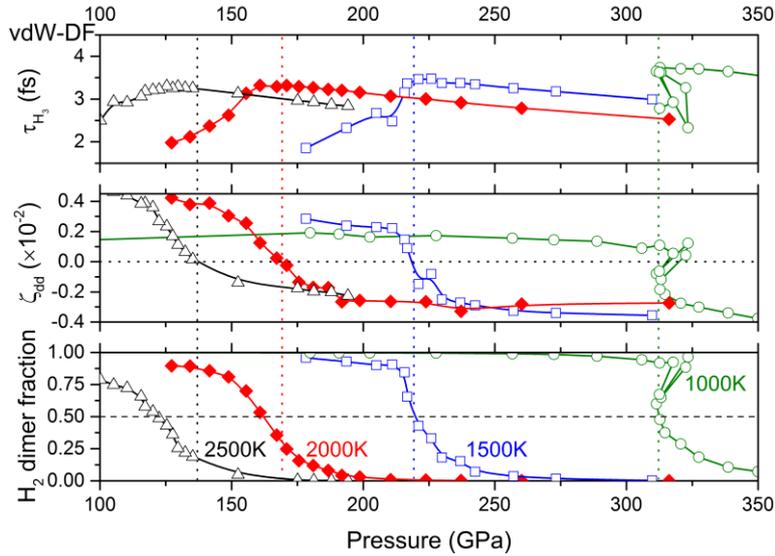

FIG. S3. Calculated fraction of stable $H_2$ molecules, dimer-dimer length covariance function $\zeta_{dd}$, and the life time of transient $H_3$ unit $\tau_{H_3}$ using vdW-DF functional as a function of pressure. Open circles—1000 K, open squares—1500 K, filled rhombus—2000 K, open triangle—2500 K.

The structural evolution of the system across the dissociation region can be further understood with the aid of Fig. S3, in which the fraction of stable dimers, the dimer-dimer length covariance $\zeta_{dd} = \langle (d_1 - \langle d \rangle)(d_2 - \langle d \rangle) \rangle$, and the lifetime of transient $H_3$ units as a function of pressure and temperature are shown. It is worth pointing out that we observed a distinct *hysteresis* at 1000 K in both PBE and vdW-DF calculations, which is an unequivocal signature of first-order LLT. Though the sharp jump in density due to the LLT has been observed previously, to the best of our knowledge, it is the first time that such a striking hysteresis is obtained in first-principles MD simulations, thanks to the ultrafine density grid employed here. There is no hysteresis beyond 1500 K, where the dissociation gradually becomes a smooth cross-over. As shown in the main text, large fluctuations in thermodynamic quantities are also encountered during long-time AIMD simulations at 1000 K. These fluctuations are also smoothed out with increased temperature.

**II. Additional results calculated with PBE and vdW-DF functional**

In the main text, typical results calculated with PBE and vdW-DF functional are presented and discussed. For completeness, additional results calculated with PBE functional are presented here. Even though the dissociation pressure and density calculated by PBE and vdW-DF functional are significantly different, the nature of the transition is similar, as shown in Figs. S4~S9.

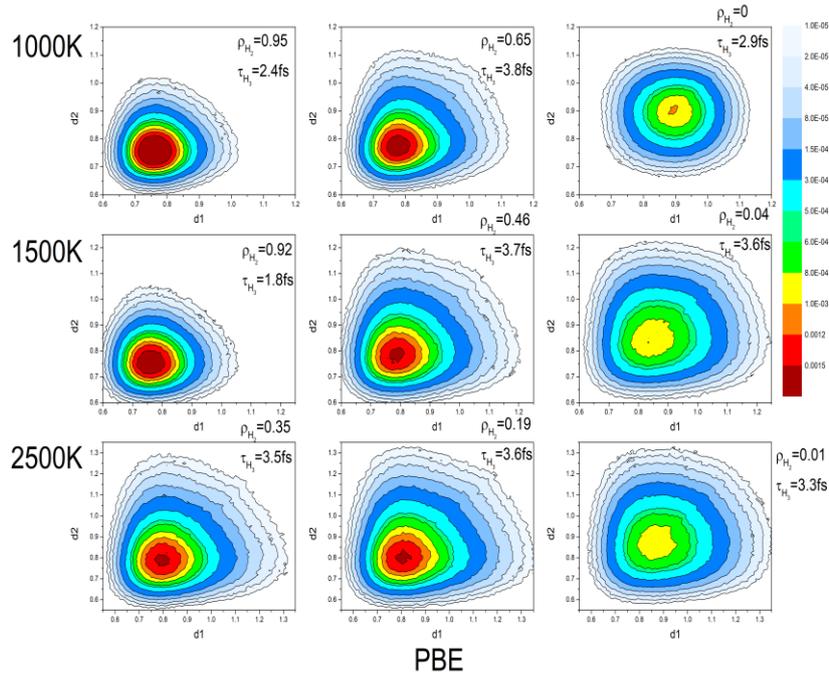

FIG. S4. DDLD in warm dense liquid hydrogen near $H_2$ dissociation calculated with PBE functional. The correlation is represented by a probability density distributed in a space spanned by the dimer length of two nearest neighboring pairs, which also include the possible $H_3$ clusters (*i.e.*, two pairs that sharing a central atom). Results computed with PBE functional at different temperatures and pressures are presented, with variations in dimer fraction $\rho_{H_2}$ and $H_3$ cluster lifetime $\tau_{H_3}$, across the dissociation region. The $H_3$ cluster has a very short life time, so any instance where three hydrogens are found adjacent is included. By contrast, the dimer fraction $\rho_{H_2}$ only includes pairs separated by less than 0.95 Å for more than 75 *fs*. The statistics for the dimer-dimer length distribution, however, is over all possible (including transient) dimers and their nearest neighboring pairs that lie within a distance of $r_{cut} = r_{NNN} + \frac{r_{NN}}{2}$ to the mass center of a given dimer, in which $r_{NN}$ is the given dimer's length and $r_{NNN}$ is the largest next nearest neighbor distance for the atoms in that dimer, for the purpose to capture the transient and short-range dimer-dimer correlation during dissociation. The same setting is used for Fig. S2, as well as Figs. S1 and S5, except that where only independent dimers are counted.

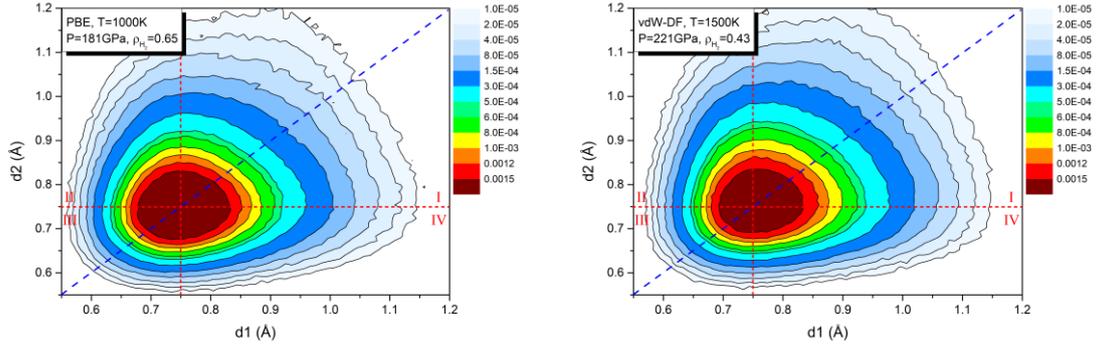

FIG. S5. DDLD in the dissociation region for independent dimers (after excluding any contribution from $H_3$ clusters). Left: PBE at 1000 K; right: vdW-DF at 1500 K. No positive correlation can be observed.

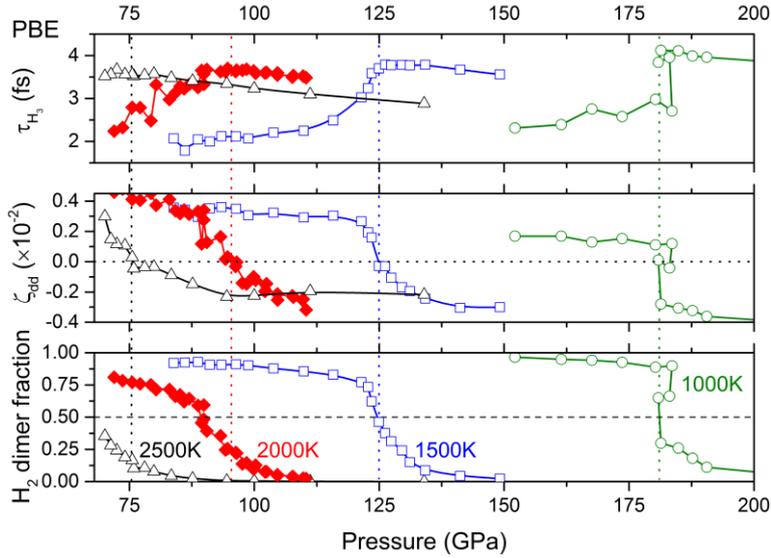

FIG. S6. The fraction of stable $H_2$ dimer, dimer-dimer length covariance function $\zeta_{dd}$, and the lifetime of transient $H_3$ unit $\tau_{H_3}$ as a function of pressure calculated by PBE. Open circles—1000 K, open squares—1500 K, filled rhombus—2000 K, open triangle—2500 K. The hysteresis at 1000 K is evident. Rare events of barrier crossing worsen the statistics quality, and manifest as noise in the data set, which is apparent in the cases of 2000 and 2500 K.

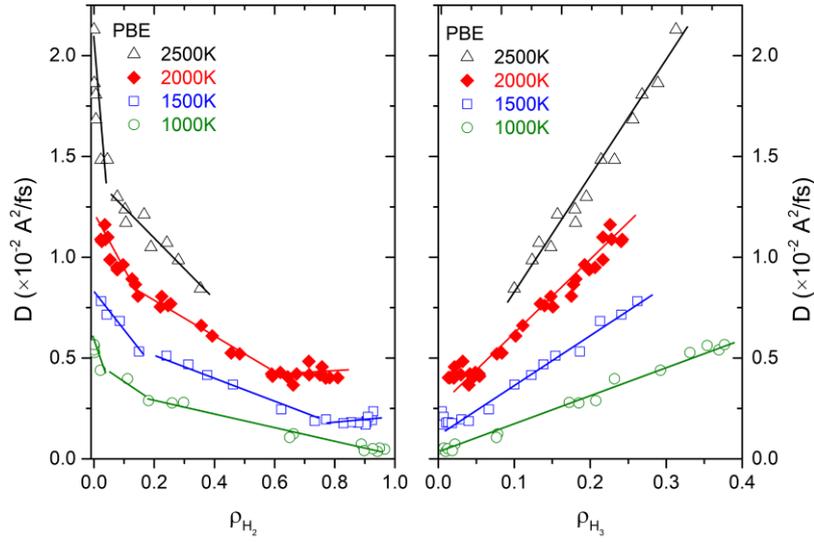

FIG. S7. Self-diffusion constant of protons in the vicinity of hydrogen dissociation as a function of stable $H_2$ molecule and transient $H_3$ unit fraction, respectively. These results were calculated by using PBE. Solid lines are guides to the eye only. No kink or jump can be observed at or near $\rho_{H_2} = 0.5$.

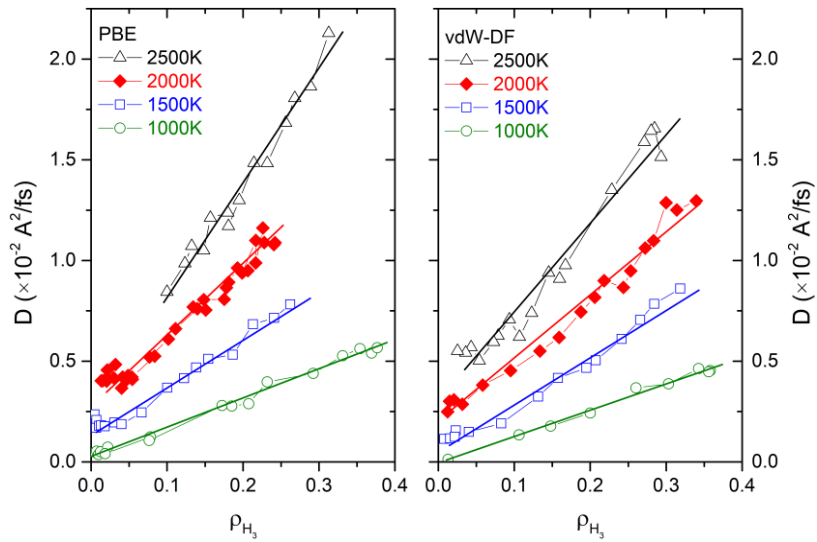

FIG. S8. Comparison of the self-diffusion constant of protons in the vicinity of hydrogen dissociation as a function of transient $H_3$ unit fraction calculated by PBE and vdW-DF functional, respectively. The vdW-DF always predicts a lower mobility within the dissociation, might be due to its predicted higher dissociation density (and thus higher pressure). Bold solid lines are guides to the eye only.

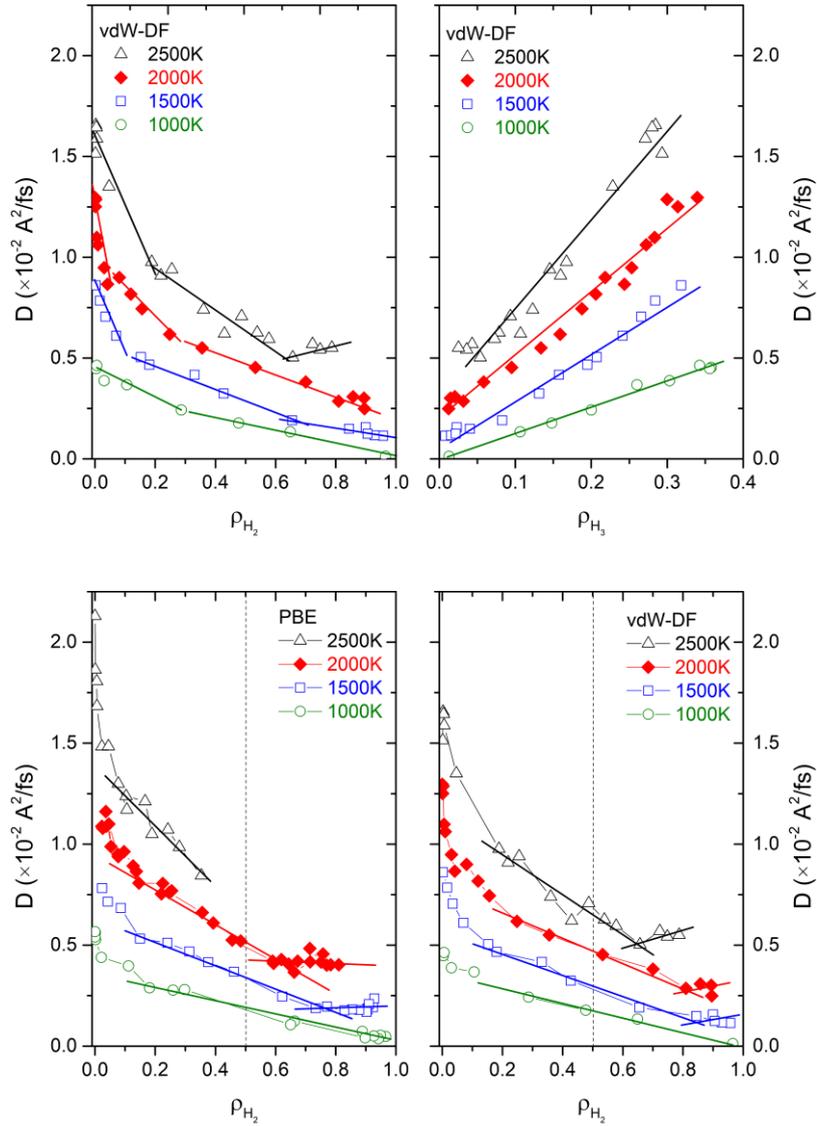

FIG. S9. Upper panel: Self-diffusion constant of protons using vdW-DF in the vicinity of dissociation as a function of stable $H_2$ dimer and transient $H_3$ unit fraction, respectively. Lower panel: Comparison of the self-diffusion constant of protons in the vicinity of hydrogen dissociation as a function of stable $H_2$ dimer fraction calculated by PBE and vdW-DF functional, respectively. There are distinct segmental regions, but no kink or jump can be observed at or near $\rho_{H_2} = 0.5$. Bold solid lines are guides to the eye only.

As can be seen in Figs. S7~S9, the proton diffusivity increases steadily when crossing the dissociation region, along with an increased pressure and density. This is counter-intuitive, since usually particle mobility will reduce as the pressure (and the density as well) increases. The rapid increase of the proton self-diffusion constant at when $\rho_{H_2} \to 0$ is due to the percolation of atomic H, revealing that proton diffusivity is dominated by fast-moving H-atoms, and being insensitive to the sharp first-order LLT. It also explains the diffusivity-$\rho_{H_3}$ relationship as shown in Figs. S7~S9. That is, the higher "$H_3$" fraction indicates more fast-moving atomic H, thus the diffusivity becomes almost linear with respect to $\rho_{H_3}$.

### III. Thermodynamic modelling of the dissociative transition

#### A. Dimer fraction variation across the dissociation region by AIMD

AIMD simulation shows that the variation of the stable $H_2$ dimer fraction across the dissociation region is smooth at high temperatures. A sharp jump was observed only when below 1500 K, as shown in Fig. S10.

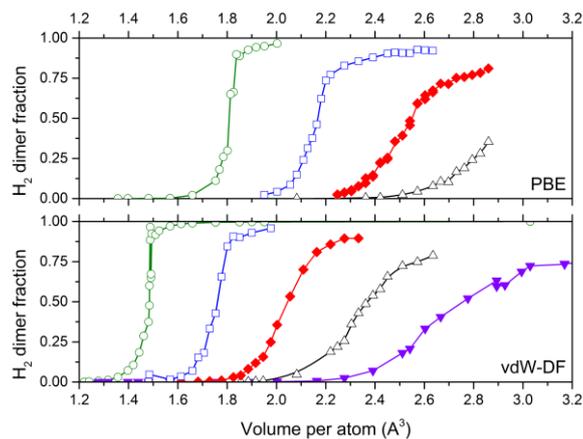

FIG. S10. The fraction of stable $H_2$ dimer as a function of the atomic volume calculated by PBE and vdW-DF, respectively. The behavior is qualitatively the same for the two functionals, except for the shift in pressure. Open circles—1000 K, open squares—1500 K, filled rhombus—2000 K, open triangle—2500 K, filled triangle—3000 K.

#### B. The non-ideal solution model

The Gibbs free energy relates to the Helmholtz free energy via

$$G(P,T) = F(V,T) + PV.$$

We approximate the $H_2$-dissociation as a two-component solvation process in a chemical picture, then we have

$$G(P,T,x) = k_B T[x\ln x + (1-x)\ln(1-x)] + xF_{H_2}(V,T) + (1-x)F_H(V,T)$$
$$+ x(1-x)J + PV \qquad (1)$$

in which $F_{H_2}$ and $F_H$ are the Helmholtz free energy component of the molecular and atomic hydrogen, respectively. The first term on the right-hand side of Eq.(1) corresponds to the ideal mixing entropy. The parameter $J$ is introduced to describe the possible interaction between the two components (*i.e.*, the coupling strength between $H_2$ and $H$), which describes non-ideal mixing. The equilibrium volume $V$ at a given pressure $P$ now becomes a function of temperature $T$ and molecular fraction $x$ via $V = \left(\frac{\partial G}{\partial P}\right)_T$. If we assume that $J$ is invariant during the dissociation process at a given pressure and temperature (namely, $\left(\frac{\delta J}{\delta x}\right)_{P,T} = 0$). The thermodynamic equilibrium condition of $\left(\frac{\delta G}{\delta x}\right)_{P,T} = 0$ then leads to

$$k_B T \ln\frac{x}{1-x} + xf_1 + f_0 = 0, \qquad (2)$$

with

$$f_1 = -\left(\frac{\delta \Delta F}{\delta x}\right)_{P,T} - 2J,$$

$$f_0 = \left[P + \left(\frac{\partial F_H}{\partial V}\right)_T\right]\left(\frac{\partial V}{\partial x}\right)_{P,T} - \Delta F + J,$$

in which $\Delta F = F_H - F_{H_2}$. This nonlinear equation could have multiple solutions, for which a formal expression can be written as

$$x = \frac{1}{e^{(xf_1+f_0)/k_B T} + 1},$$

and can be solved self-consistently.

At the critical point of LLT, there should be just a single solution for $x$ that minimizes $G$, and must satisfy $\left(\frac{\delta^2 G}{\delta x^2}\right)_{P,T} = 0$. The latter condition results in (after ignoring higher-order contributions)

$$\frac{k_B T}{x(1-x)} + 2f_1 + 2J = 0. \tag{3}$$

The solutions are

$$2x = 1 \pm \sqrt{1 - \frac{2k_B T}{J + \left(\frac{\delta \Delta F}{\delta x}\right)_{P,T}}}.$$

For the dissociation of diatomic molecules, the critical point must occur at $x = 1/2$, then we have

$$2k_B T_c = J + \left(\frac{\delta \Delta F}{\delta x}\right)_{P,T},$$

$$\Delta F|_{P=P_c, T=T_c} = \frac{J}{2} - k_B T_c + \left[P_c + \left(\frac{\partial F_H}{\partial V}\right)_{T_c}\right]\left(\frac{\partial V}{\partial x}\right)_{P_c, T_c}.$$

The latter relation is after Eq.(2), which constrains the Helmholtz free energy difference between $F_{H_2}$ and $F_H$ at the critical point ($P = P_c, T = T_c$).

When $2k_B T < J + \left(\frac{\delta \Delta F}{\delta x}\right)_{P,T}$, Eq.(3) has two distinct solutions, which correspond to the lower-bound and upper-bound of the instability region (*i.e.*, the spinodal curves) for the two-phase coexistence of the first-order transition, respectively. The corresponding $x$ for the stable H-rich and $H_2$-rich phases are given by the solutions of Eq.(2) (*i.e.*, the coexistence curves). When $2k_B T > J + \left(\frac{\delta \Delta F}{\delta x}\right)_{P,T}$, Eq.(3) has no real solution, and the single solution of Eq.(2) determines the equilibrium dimer fraction for the dissociative cross-over (continuous) transition. This crossover can be approximately modeled by neglecting the H-$H_2$ coupling and the variation of $\Delta F$ during dissociation, which is equivalent to setting $T_c \to 0$. Considering that the first-order feature (*i.e.*, the hysteresis) in LLT of dense liquid hydrogen is very weak, this approximate treatment is plausible. Furthermore, the volume jump of the first-order transition in fact can be accurately captured by this approximation, since at low temperatures, the continuous variation of thermodynamic quantities of this model becomes sharp, leading to a pseudo-transition that resembling a first-order "jump" [63,64].

**C. Pseudo-transition model (PTM)**

The theory of pseudo-transition and its application in nonstoichiometric compounds, as well as the induced thermodynamic anomalies and the implications in physics, had been elaborated in Refs. [63,64]. Here we adapted the theory for diatomic molecular dissociation.

Assuming the dissociation can be viewed as a physical process of an *isolated* two-level system that is governed by the energy difference $\Delta E$ between these two levels, which we called the binding free energy for a diatomic molecule case, Eq.(1) reduces to the Helmholtz free energy per particle as expressed as

$$F = k_B T[x \ln x + (1-x)\ln(1-x)] - x\Delta E.$$

Here $\Delta E$ is defined to be a positive quantity. Compared with Eq.(1) above we are incorporating interaction terms $J$ and the PV terms into $\Delta E$. Minimizing $F$ with respect to the molecular fraction $x$ gives

$$x = \frac{1}{e^{-\Delta E/k_B T} + 1},$$

which is exactly what we used to model the dissociation of hydrogen molecules in the main text with $\rho_{H_2} \equiv x$ and $\Delta E/k_B = a - bP - cT - dTP$, see Figs. S11 and S12. It should be noted that since this two-level system is a mathematical modelling of the thermal excitation of a dissociation process, only the entropy of level-occupation is explicitly considered. All other entropy and internal energy contributions have been put into $\Delta E$, which also includes other terms of higher orders in pressure and temperature. In this model, the dissociation phase line on the P-T plane is described by an equation of

$$P = \frac{a - (c + \ln y)T}{b + d \cdot T}$$

with $y = 1$ for the dissociation middle line (*i.e.*, the Widom line), and $y = \frac{7}{3}$ and $\frac{3}{7}$ for the 3/7 lines, respectively (see Fig. S12). For a given temperature $T$, the width of the dissociation region bracketed by the 3/7 lines is thus

$$\Delta P = \frac{2T \ln(7/3)}{b + d \cdot T}$$

which approaches zero when $T \to 0$, *i.e.*, the presumed critical point.

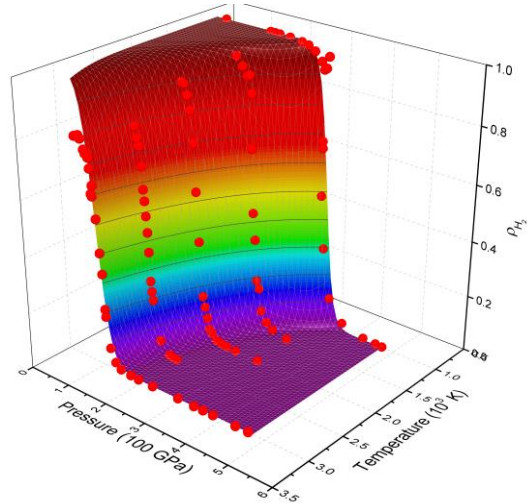

FIG. S11. Variation of the stable $H_2$ dimer fraction with pressure and temperature obtained with vdW-DF: solid points—DFT results, surface—fitting to the DFT data using an equation of (pseudo-transition model) $\rho_{H_2} = \left(\exp\left(\frac{-a+bP+cT+dTP}{T}\right) + 1\right)^{-1}$, in which the temperature T and pressure P are in a unit of 1000 K and 100 GPa, respectively. The fitting parameters are *a*=36.4962, *b*=3.34851, *c*=5.73474, and *d*=6.10592. One can include higher order terms in P and T to account for the anharmonic vibrational and other entropy contributions, but within our studied temperature and pressure range, these terms can be ignored.

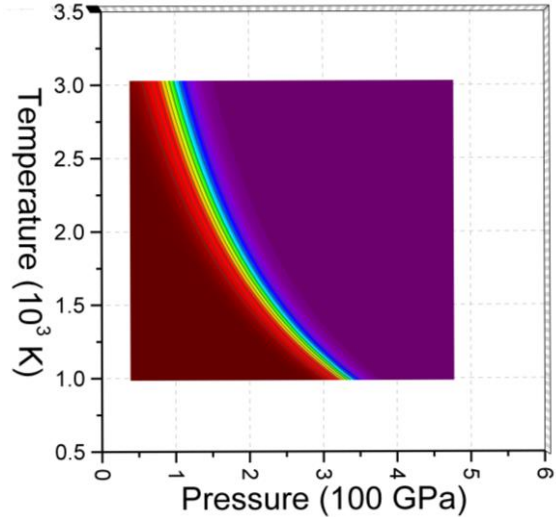

FIG. S12. Dissociation boundary of warm dense liquid hydrogen predicted by a pseudo-transition model. The same as Fig. S11 but projected onto the T-P plane, in which the spread region of $H_2$ dissociation at high temperatures is shown. This figure is also presented as an inset in Fig. 6(a) of the main text.

### D. A variant of the PTM

Alternatively, one can take the dissociation entropy of $H_2 \rightarrow 2H$ into account, explicitly recognizing the increase in number of particles, and rewriting the Helmholtz free energy in a Sackur-Tetrode based form of

$$\frac{F}{k_B T} = -N_m \ln\left(\frac{Ve^{5/2}}{\lambda_m^3}\right) + N_m \ln N_m - (N - 2N_m)\ln\left(\frac{Ve^{5/2}}{\lambda_A^3}\right)$$

$$+ (N - 2N_m)\ln(N - 2N_m) - N_m \frac{\Delta E}{k_B T},$$

in which $\lambda$ is the thermal de Broglie wavelength, $N$ is the total number of atoms, of which $2N_m$ are bounded into molecules. Minimizing this function with respect to $N_m$ leads to

$$\ln\left(\frac{Ve^{5/2}\lambda_m^3}{\lambda_A^6}\right) - \ln\frac{(N - 2N_m)^2}{N_m} = \frac{\Delta E}{k_B T} + 1$$

or

$$\frac{(N-2N_m)^2}{N_m} = \frac{Ve^{3/2}\lambda_m^3}{\lambda_A^6} e^{\frac{-\Delta E}{k_B T}} \equiv BN.$$

This gives a dimer fraction of

$$x \equiv \frac{2N_m}{N} = 1 - \frac{B}{4}\left(\sqrt{1+\frac{8}{B}} - 1\right)$$

with parameter $B = \frac{v(e/2)^{3/2}}{\lambda_A^3} e^{\frac{-\Delta E}{k_B T}}$, the atomic volume $v = V/N$, and the atomic thermal de Broglie wavelength $\lambda_A = h/\sqrt{2\pi m k_B T}$.

The phase line equation for dissociation on the T-v plane given by this model is

$$\frac{3k_B T}{2} + k_B T \ln\left[\frac{(\pi m k_B)^{\frac{3}{2}}}{yh^3}\right] + k_B T \ln(vT^{\frac{3}{2}}) = \Delta E(v, T)$$

with $y = 1$ for the dissociation middle line, and $y = \frac{49}{15}$ and $\frac{9}{35}$ for the 3/7 lines, respectively.

If we simply assume $\Delta E/k_B \approx a - \frac{b}{v} - cT - d\frac{T}{v}$, the same data set as shown in Fig. S11 can be fitted to this variant PTM, with the fitting parameters $a$=52.97, $b$=46.83, $c$=-5.63, and $d$=34.73, in which the temperature is in units of 1000 K and the atomic volume in Å$^3$. One should note, however, that the parameter "$a$" here does not correspond to the binding energy of a free diatomic molecule, because the functional form of $\Delta E$ we employed here is too simple: it is applicable only in the vicinity of the H$_2$ dissociation, and obviously becomes invalid at lower pressures. This difficulty can be solved if a more physically sensible binding free energy as a function of temperature and atomic volume is employed.

**IV. Isotherms across the dissociation region**

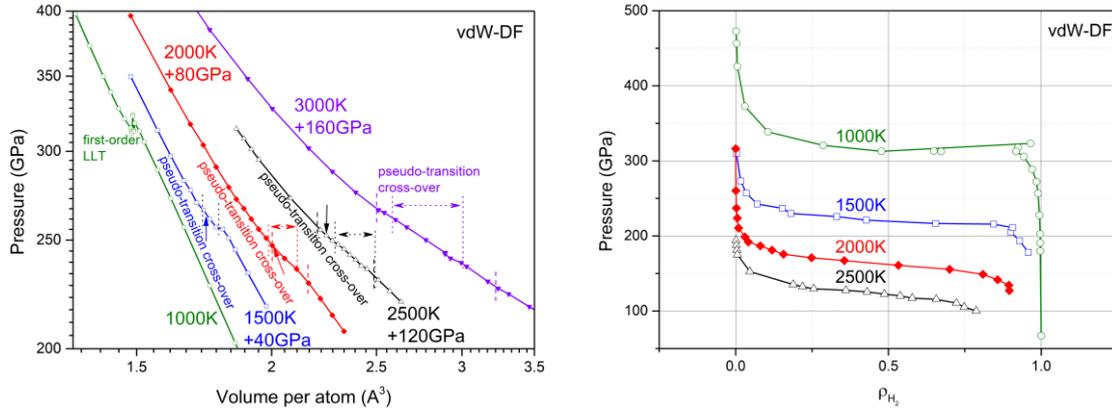

FIG. S13. Left panel: Isothermal pressure-volume curves of warm dense liquid hydrogen across the dissociation region predicted by vdW-DF. Curves for 1500 K, 2000 K, 2500 K, and 3000 K are up-shifted by 40 GPa, 80 GPa, 120 GPa, and 160 GPa, respectively. An evident first-order transition (LLT) with hysteresis is obtained only at low temperatures. The short vertical dashed lines mark the transition/cross-over region determined by visual judgement of the changes in the curve slope. The short vertical dotted lines denote the region with a stable $H_2$ fraction between 0.3 and 0.7, respectively. The arrows indicate the condition where the transient $H_3$ unit attains the largest lifetime. The dimer-dimer length covariance function also changes its sign at the same point. The same setting was used for Fig. 2 in the main text. Right panel: the same data set as the left panel but without pressure-shifting, it displays the variation of pressure along with the $H_2$ dimer fraction.

The isothermal compression curves across the dissociation region are plotted in Fig. 2 of the main text and Fig. S13 for the PBE and vdW-DF functional, respectively. The volume collapse and hysteresis are evident for the case of 1000 K. It becomes less perceptible at 1500 K. At higher temperatures of 2000 and 2500 K, the curves are broad cross-overs, and it is ambiguous to associate them with a first-order transition. Even so, we still estimated a "transition line" by taking the middle point of the slope change of the curve. This line extends beyond the critical point, as shown in Fig. 6 of the main text which is in good agreement with previous estimates, but does not always correspond to the mid-point of the dissociation. It is especially the case for the PBE results at 2500 K.

It is worth emphasizing that, with these isotherms on the P-V plane, it is not possible to construct van der Waals loops at around the volume jumps like the well-known liquid-vapor transition (see Fig. 2 of the main text). It is thus impossible to locate the critical point by tracing the successive variation of these isotherms. In fact, there is no well-defined phase separation region on the P-V plane here (otherwise the isotherms with a lower temperature would exhibit a kink at a lower pressure than those with a higher temperature). That is, at the given density that corresponds to the critical point $(P_c, T_c)$, by decreasing the temperature from $T > T_c$ to $T < T_c$ via $T = T_c$ along the isochoric line, we cannot have a spontaneous symmetry breaking that drives the system transform from a homogeneous single phase into a heterogeneous two-phase coexistence. This phenomenon is due to the fact that the order parameter of the dissociative transition is not the density difference. Instead, as shown in previous section, the proper order parameter should be the dimer fraction. Therefore, the two-phase coexistence occurs in the P-$\rho_{H2}$ or T-$\rho_{H2}$ space (see the right panel in Fig. S13), rather than the P-V space. That is why in the real space the $H_2$/H interface is volatile.

**V. Transient clusters within the dissociation region**

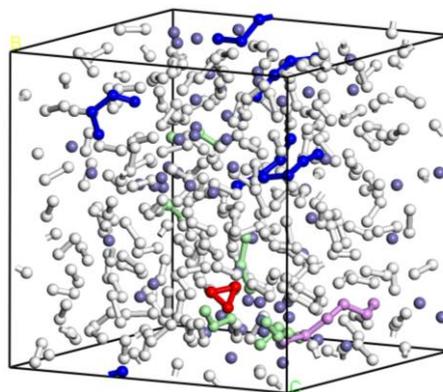

FIG. S14. A snapshot of atomic configuration taken from one AIMD simulation of PBE at 2000 K. The pressure is about 96 GPa, with an atomic volume of 2.42 Å$^3$ and an $H_2$ dimer fraction of 0.22. This condition is in the vicinity where the dimer-dimer length covariance function changes sign. The atomic H (gray), diatomic $H_2$ (white), transient $H_3$ units (red for the acute

internal angle and green otherwise), as well as other longer transient chains of hydrogen, can be seen. There is no sign of phase separation between atomic H and molecular $H_2$ entities.

Within the dissociation region, extensive AIMD simulations identified many large transient clusters (Fig. S14), often associated with rebonding events. In addition to the stable $H_2$ and atomic H, the most common entity is $H_3$, which however, is short lived (Figs. S3 and S6). The consequence of formation these transient clusters is that they lead to inter- and intra-entity charge fluctuation/sloshing (Figs. S15 and S16. Note that they are not always positively charged as previously thought, but tend to be neutral). This leads to polarization and polarizability, and gives rise to novel optical properties. For example, it will activate prohibited infrared or Raman modes, and generate a strong and noisy background.

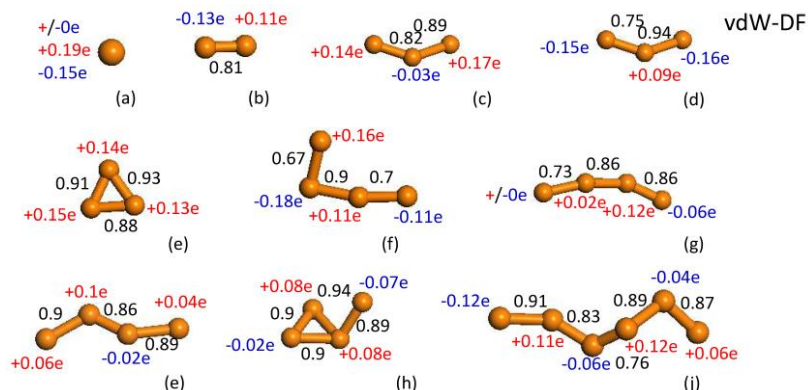

FIG. S15. Illustration of some representative types of hydrogen clusters, together with their bond length and Bader charge, calculated with vdW-DF at 2500 K. The pressure is about 122 GPa, with an atomic volume of 2.36 Å$^3$, and an $H_2$ dimer fraction of 0.49.

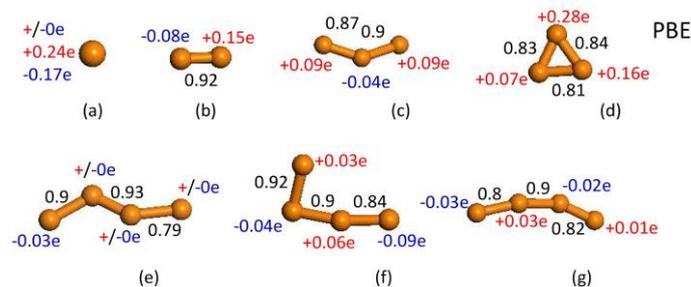

FIG. S16. Illustration of some representative types of hydrogen clusters, together with their bond length and Bader charge, calculated with PBE at 2000 K. The pressure is about 96 GPa, with an atomic volume of 2.42 Å$^3$, and an $H_2$ dimer fraction of 0.22.

## VI. Angular distribution function (ADF) of $H_3$ cluster

The angular distribution function (ADF) of $H_3$ clusters are evaluated and plotted in Figs. S18 and S19, from which one can see that the most frequent geometry is that with an angle of $130°$. Linear or triangular trimers are very rare. Angles below $60°$ or above $180°$ are impossible by definition. Another indication from these ADFs is that PBE has stronger local structural features than vdW-DF, and is more pronounced at low temperature and high density where angular orientation is more correlated. This observation is consistent with the previous simulation that concluded compression restricts the rotation and vibration of $H_2$, and enhances the *local* structure correlations [67]. This angular correlation at low-T and high-P changes drastically during $H_2$ dissociation, and should be the physical origin of the occurrence of the first-order LLT.

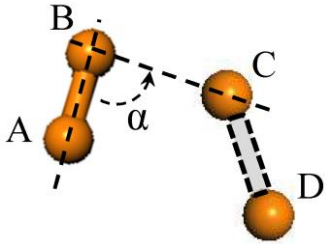

FIG. S17. Schematic demonstrating the identification of structural features of warm dense liquid hydrogen. Particle C is the nearest neighbor to the dimer AB within a cutoff radius (as defined in Fig. S4). D is the nearest neighbor to C, and CD might form a dimer or not, depending on their separation. The dimer-dimer length covariance (DDLC) function is calculated as $<r_{AB}r_{CD} - r_2^2>$, where $r_2$ is the statistically averaged dimer length. When computing the DDLC, configurations in which particle C forms a bond to particle B or A are excluded. In the case where ABC forms an $H_3$ cluster, the bond angle $\alpha$ is evaluated and shown in Figs. S18 and S19.

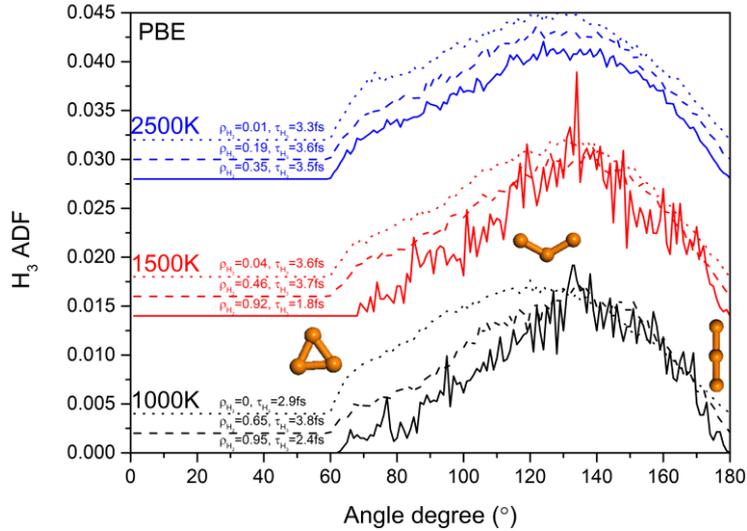

FIG. S18. Angular distribution function (ADF) of the bond angle $\alpha$ in $H_3$ clusters calculated with PBE. The curves are relatively shifted for clarity. The most common angle is around 130°. The regular triangle geometry is more frequent at relatively high temperature and with a low $H_2$ fraction.

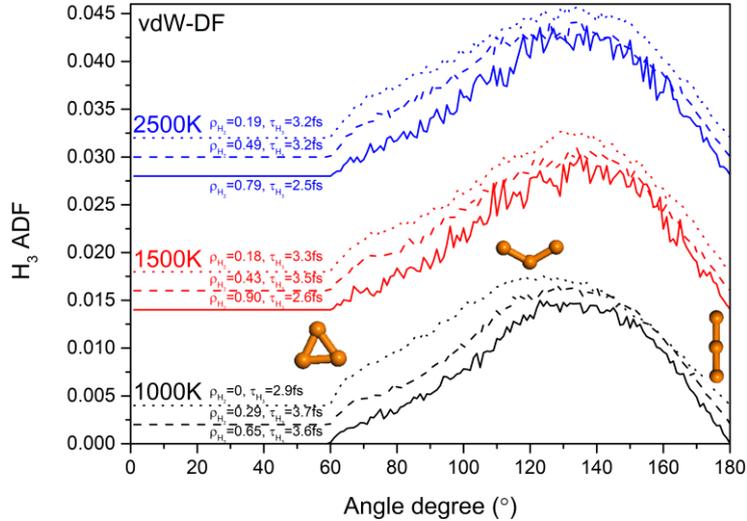

FIG. S19. The ADF of the bond angle $\alpha$ in $H_3$ clusters calculated with vdW-DF. The curves are shifted for clarity. The most frequently observed angle is around 130°. Note the small but finite probability of the geometry near the equilateral triangle, and the almost zero probability for a linear configuration. Angles below 60 degrees are impossible by definition. A fluctuation in the angular geometry is prominent when $H_2$ fraction is high. The ADF becomes smooth when $H_2$ fraction less than 0.5. These ADFs not only reflect the local structure correlation and the charge neutrality of the whole system, but also indicate a possible local charge fluctuation/sloshing.

## VII. Mixing of atomic and molecular hydrogen within the dissociation region

Direct AIMD simulations showed that there is no distinct phase separation of $H_2$/H within the dissociation region, even below $T_c$. However, the simulation cannot rule out a "slow" phase separation—which for AIMD means nanosecond timescale or above! In order to check this, we start the simulation from a two-phase coexistence configuration that is *already segregated* and has a clear interface. In order to avoid the "spacing problem" that could be introduced between the two initial subsystems or phases, the initial configuration was prepared by first heating the system up to a high enough temperature to fully dissociate $H_2$. Then half of the atoms were fixed, and the other half were cooled down to a low enough temperature to form bonded $H_2$ molecules. This treatment created a physical two-phase ($H_2$/H) interface without artificial gap. Subsequently, each of the two subsystems was fully equilibrated before they were thermodynamically combined. The pressure was evaluated for the whole system. This will introduce nonzero deviatoric stress in *NVT* ensemble if there is a density jump. To remove the deviatoric stress, a supercell-shape optimization was then performed.

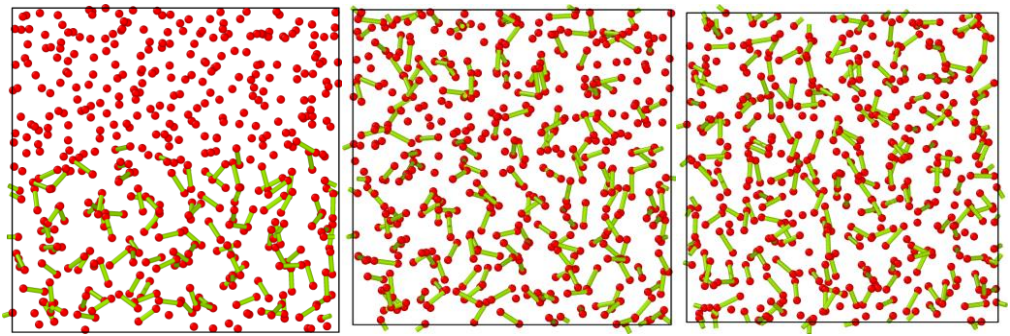

FIG. S20. Mixing of atomic H and molecular $H_2$ within the dissociation region at 125 GPa and 1500 K, calculated by AIMD with PBE functional. Left panel: initial two-phase coexistence with a clear interface; middle panel: after 5 *fs* equilibrating; right panel: after 950 *fs* equilibrating. The bond length cutoff criterion is set as 0.9 Å.

Starting from this initial configuration, at thermodynamic conditions below and above $T_c$ and well within the dissociation region, further AIMD simulations were performed. The results reveal a rapid interconversion of H and $H_2$, as shown in Figs. S20 and S21. The initial interface becomes indiscernible within 5 femtoseconds. There are three main features in this $H_2$/H mixing: (1) the mixing is not driven by diffusion of H into the $H_2$ region or *vice versa*, rather it is by formation of $H_2$ dimers within the H region, and/or by dissociating into atomic H in the $H_2$ region; (2) some $H_2$ dimers have a very short lifetime, there is an ongoing process of bonding, dissociating, diffusing, and then re-bonding; (3) at the picosecond timescale, individual atomic motion is homogeneous and isotropic across the simulation supercell—all atoms spend some time as part of molecules and some time dissociated. An animation of the AIMD trajectory of proton diffusion that clearly reveals these features is available upon request.

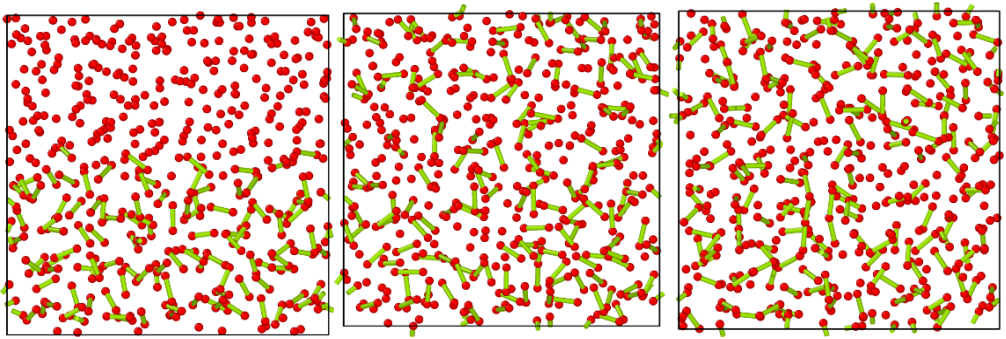

FIG. S21. Mixing of atomic H and molecular $H_2$ within the dissociation region at 181GPa and 1000 K, calculated by AIMD with PBE functional. Left panel: initial two-phase coexistence with a clear interface; middle panel: after 5 *fs* equilibrating; right panel: after 150 *fs* equilibrating.

**VIII. Effective pair potential in the vicinity of the dissociation**

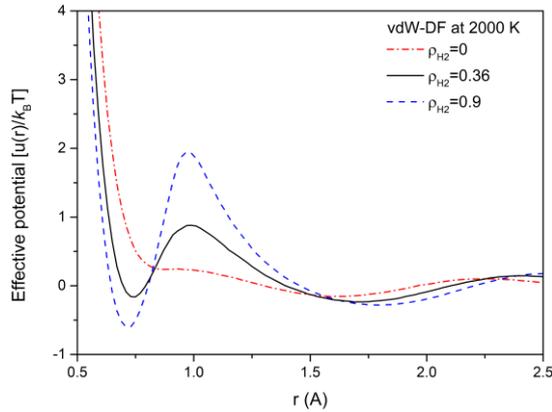

FIG. S22. Variation of the effective pairwise potential with different $H_2$ dimer fraction at 2000 K. This potential-of-mean-force is obtained by the Boltzmann inverse method from the calculated radial distribution functions. The relative broad feature of the first attractive trap for a dimer fraction of ~0.36 arises from the existence of molecules, with the strong repulsion around 1 Å representing the rarity of molecular dissociation and the intermolecular repulsion. At higher pressures, where $H_2$ is fully dissociated, the absence of a distinct potential barrier means that dimers or other clusters are no more likely than in a hard-sphere liquid.

The short lifetime of $H_3$ unit, as well as other large clusters, can be perceived from the effective pair potential between hydrogen atoms in the dissociation region (see Fig. S22): the relatively flat potential allows large clusters to form, but their "binding" is very weak, and they can be destroyed by thermal noise easily.

## IX. Thermodynamic fluctuations near the dissociation region

Thermal fluctuations encode information of the underlying potential energy landscape, thus reflecting the interaction and correlation among particles in a system. This is especially important near a phase transition. For example, the thermodynamic fluctuation becomes very large when approaching the critical point (CP) of a second-order phase transition, due to the presence of strong and long-ranged correlations, which mirrors the divergence of thermodynamic quantities at the critical point. In this regard, thermodynamic fluctuation is a useful criterion to locate $T_c$. For this purpose, we evaluated the thermal fluctuations of internal energy and pressure along the isotherms across the dissociation region. It is necessary to point out that, due to the small scale of currently accessible AIMD simulations, the quantity obtained from this fluctuation analysis is too noisy to evaluate critical exponents, nonetheless, an accurate estimate of $T_c$ or its absence can be reliably made.

The fluctuation information was extracted from *NVT* AIMD simulations that are fully equilibrated for about 1.5 *ps*, with statistics collected over another 1.5 *ps*. To reduce noise, we only evaluated contributions that can be completely determined by potential energy. Thermal contributions, *i.e.*, the kinetic energy part of $C_v = 1.5\ k_B$ and the thermal pressure of $P = Nk_B T/V$, are added later.

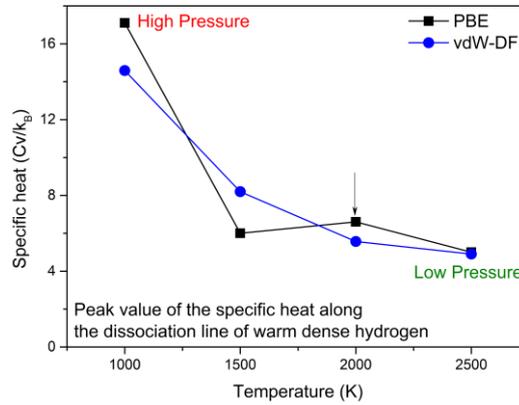

FIG. S23. Variation of the peak value of the specific heat (along the dissociation line) from low temperature to high temperature (from the high-pressure end to the low-pressure end, as well). The arrow indicates the previously assigned $T_c$ from Knudsen *et al*. [27].

Figure 4 in the main text illustrates the constant volume specific heat per atom extracted from the energy fluctuations by $k_B T^2 C_v = \langle (E - \langle E \rangle)^2 \rangle$. Here $\langle \cdots \rangle$ denotes ensemble average. Though the data are noisy, some qualitative conclusions can be drawn: *(1) the specific heat has a distinct maximum in the dissociation region.* At 1000 K, our measurement on a finite-sized system is consistent with a divergent specific heat and first order transition. At higher temperatures, it is not. *(2) The magnitude of the peak decreases along the dissociation line* from low-T high-P end to the high-T low-P end. This reduction reflects weaker correlations at high-T low-P conditions. *(3) The peak broadens along the dissociation line* towards the high-T low-P end, which implies the smearing of the dissociation boundary. *(4) These peaks do not originate in pseudo transition*, because they have very different shape [63,64]. *(5) These conclusions do not depend on the exchange-correlation functional*. Both PBE and vdW-DF predicted the very same features.

The variation of the specific heat along the dissociation line is plotted in Fig. S23. There is a sharp jump between 1000 and 1500 K, which indicates that $T_c$ should be bracketed by these two temperatures. The pressure fluctuation $\Delta P = \sqrt{\langle (P - \langle P \rangle)^2 \rangle}$ has similar features, as shown in Fig. S24.

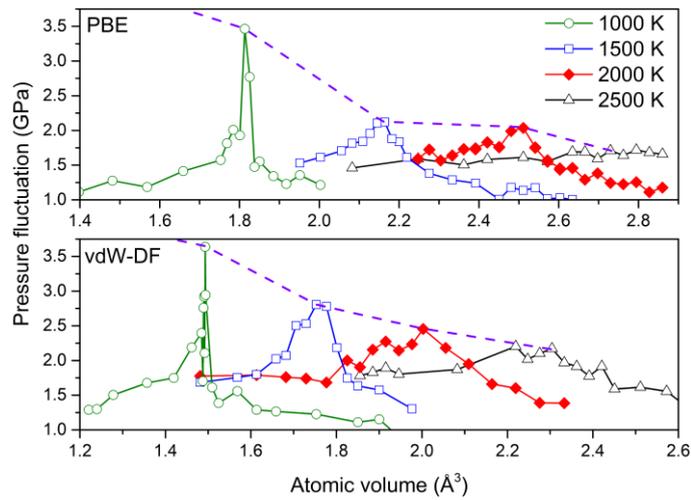

FIG. S24. Pressure fluctuation extracted from the AIMD simulations at several temperatures calculated with PBE and vdW-DF, respectively. The dashed lines that connect the peaks are guides to the eye only.

## X. Isotope effect on the location of H$_2$ and D$_2$ dissociation

As mentioned in the main text, there might be an isotope effect on the location of dissociation. A recent reflectance measurement also indicated a similar effect on the metallization of liquid hydrogen and deuterium, a further confirmation of an early dynamic compression experimental observations [72]. Our AIMD simulations did not include nuclear quantum effects (NQE), thus might overestimate the dissociation temperature and pressure. On the other hand, the PBE functional tends to underestimate the dissociation transition pressure. This "error cancellation" may give a result that is very close to the experimental one.

As shown in Fig. S25 that is a simplified version of Fig. 6 in the main text but including the recent DAC experimental data on hydrogen and deuterium [72], vdW-DF functional fits data for deuterium, whereas the cancellation in PBE functional for hydrogen still has some residual. Though the PBE results are in good agreement with the experimental data obtained by the "temperature *vs* laser heating power" curves (Fig. 6 in the main text), there is a systematic offset with respect to the latest reflectance experiment and CEIMC data. The residual error is estimated to be about 25 GPa in pressure and 200~400 K in temperature, as shown by the shifted PBE dissociation boundary in Fig. S25.

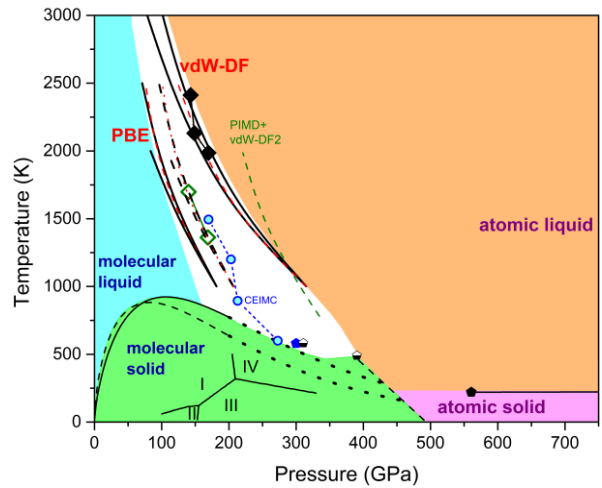

FIG. S25. Phase diagram of warm dense hydrogen, a simplified version of Fig. 6 in the main text, to illustrate the isotope effect in dissociation location. The recent reflectance experimental data (rhombus: open points for hydrogen and filled ones for deuterium, respectively) [72] are included for comparison. The shifted PBE dissociation boundary (dashed lines, shifted by 25 GPa) matches the experimental and CEIMC data for hydrogen very well.

**XI. Assessment of the viscosity in the vicinity of dissociation region**

The viscosity of warm dense hydrogen in the vicinity of the dissociation transition, which is a measure of the resistance of the liquid to gradual deformation by shear stress or tensile stress, might have important implications in Jovian planets. This transport quantity can be calculated using the Stokes–Einstein equation $\eta = \frac{k_B T}{6\pi D r}$, in which $D$ is the diffusion constant, $r$ is the radius of the spherical particle. With the assumption that the dimer can be effectively treated as a sphere, and a further approximation that the radius of a dimer relates to the radius of atomic hydrogen by $r_{H_2} = 2xr_H$, we can approximate the viscosity of the H$_2$/H liquid by the diffusivity of virtual particles with a radius of $r = \rho_{H_2} r_{H_2} + (1 - \rho_{H_2}) r_H = (1 + (2x - 1)\rho_{H_2}) r_H$. Taking the simplest approximation of $x = 1$, we obtained the viscosity from the proton diffusion constant and H$_2$ dimer fraction across the dissociation, as shown in Fig. 5 of the main text.

It is evident that the viscosity drops as a function of pressure in the vicinity of dissociation, and the drop is the most prominent in the first-order LLT region. One should note that usually viscosity increases with increasing pressure. The observed anomalous plummet as shown in Fig. 5 not only originates in the H$_2$ dissociation (the small and light atomic H drifts faster than the big and heavy H$_2$ molecule), but also relates to the fact that the studied dissociation region is very close to the anomalous melting curve of dense hydrogen that continuously decreases with increasing pressure [3]. When far away from this anomalous region, the proton viscosity returns to normal behavior, increasing with pressure.

## XII. Impact to the convection cell in cold Jovian planets

Our results showed that the $T_c$ of the first-order LLT is too low to be relevant to known Jovian planets. The wide H$_2$/H dissociative cross-over layer, however, exists far beyond $T_c$, and its strong thermodynamic anomalies are highly relevant to the internal condition of cold giant gas planets. The presence of this anomalous layer with a negative thermal expansivity will disturb the interior convection cell, and cut it into two parts, due to the instability induced by heat exchange and negative expansivity, as schematic in Fig. S26.

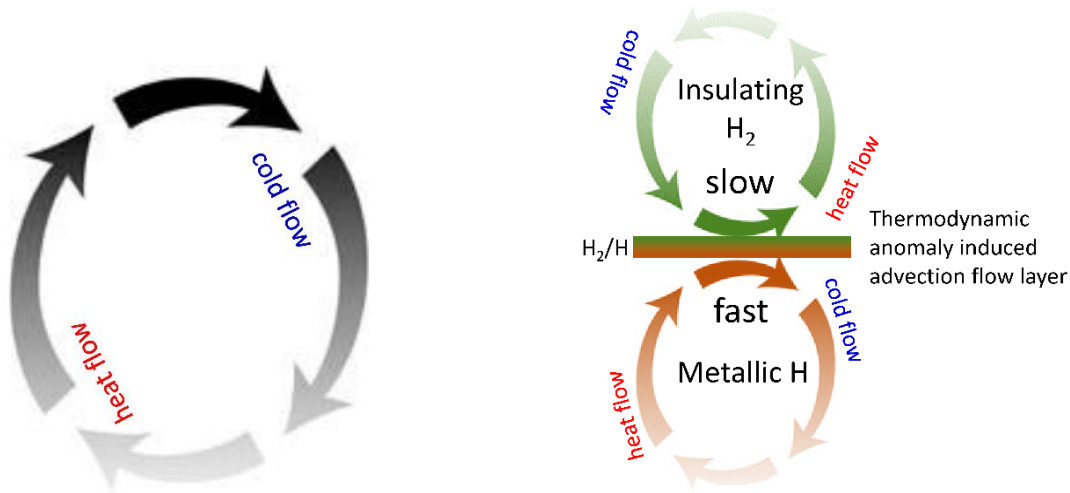

FIG. S26. Schematics that illustrate how anomalous dissociative layer modifies the convection cell in Jovian planets. Left: the convection cell in a normal fluid that usually has an "O" shape; Right: the anomalous $H_2$/H layer with a negative thermal expansivity induces an instability to the convection flow cycle, leading to an "8" shape and an advection flow in between.

## XIII. Finite size effects

Some DFT [49], VMC-MD [46,69] and CEIMC [35,50] simulations employed a small supercell that has about 100 or less atoms to simulate the liquid dense hydrogen. Here we employed a much bigger supercell with 500 H. A too-small supercell could have finite size effects, and cause a spurious first-order transition, making $T_c$ too high. From our AIMD simulations, we find that the liquid is highly structured—there are one intramolecular and up to five intermolecular peaks. With 64 hydrogens it is simply not possible to describe this structure: there are artefacts from the periodic boundary conditions between the first and second shell of molecular neighbors.

In order to check this further, we carried out other simulations with 54, 432, and 3456 atoms, respectively. As shown in Fig. S27, the RDF of the 54-atom simulation is overstructured. The delta functions in this simulation represent the lattice vectors of the supercell. One can see that in the 54-atom cell, beyond the second neighbor shell, the liquid structure is unrelated to its converged value, being entirely determined by the periodic boundary conditions. On the other hand, the 432-atom simulation (slightly smaller than our production runs that has 500 atoms) agrees with the very large system to the fourth shell, beyond where the structure loses order and becomes homogeneous. Finite size effects cause differences to appear at all separations, *e.g.*, the 54-atom cell has a significantly sharper first peak. Previous works involving such small supercells are not based on a correct description of the liquid structure [49][46,69][35,50].

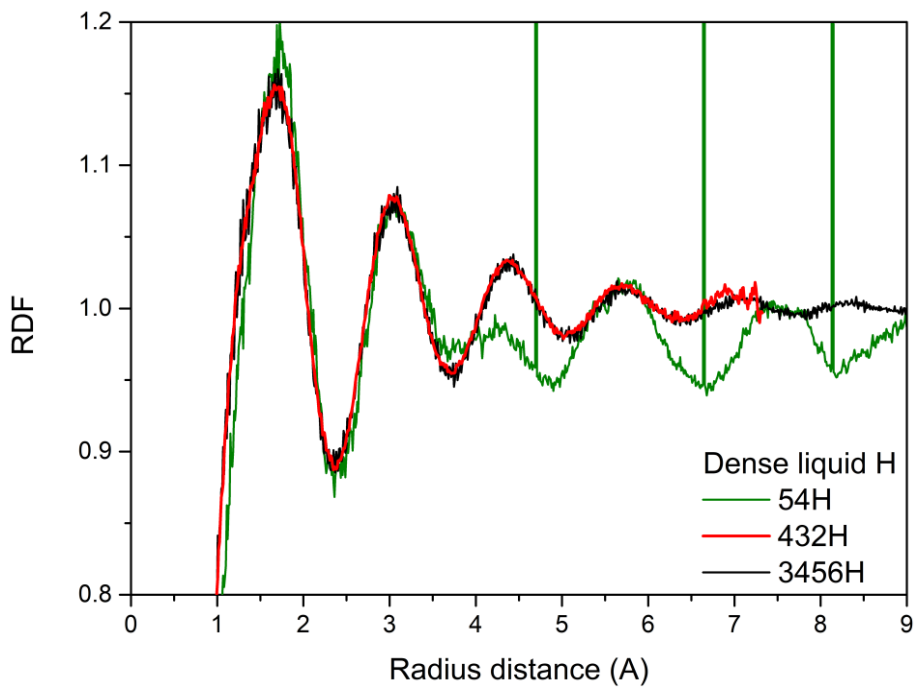

FIG. S27. Radial distribution functions for dense liquid hydrogen calculated with PBE and *NVT* simulations in a supercell with 54, 432 and 3456 atoms with periodic boundary conditions at 1000 K and a volume of 1.9226 Å$^3$/atom, respectively.

**Supplementary References**